\documentclass{iopjournal}

\usepackage{amsmath,amssymb}
\usepackage{dcolumn}
\usepackage{bm}
\usepackage{url}

\graphicspath{{../}}
\newenvironment{widetext}{}{}

\begin{document}

\articletype{Paper}

\title{Equivalence between the Axion Invariant and the $S_4$ Symmetry Indicator}

\author{Mengyao Zhang}

\affil{School of Physics, Peking University, Beijing 100871, China}

\email{Contact author: mengyao@stu.pku.edu.cn}

\date{\today}

\keywords{topological insulator, axion insulator, symmetry indicator, $S_4$ rotoinversion, magnetoelectric polarizability, Chern--Simons invariant}

\begin{abstract}
The equivalence between the axion invariant and the $S_4$ symmetry indicator is established for three-dimensional $S_4$-symmetric axion insulators with vanishing three-dimensional Chern numbers.
Starting from the Chern-Simons expression for the magnetoelectric polarizability, $2P_3=\theta/\pi$ is rewritten in terms of the $S_4$ sewing matrix. 
After stable reduction to determinant-one two-band blocks, the invariant is expressed as the degree of a map from the Brillouin zone to $SU(2)$. 
The degree modulo two is then evaluated from the $S_4$ eigenvalues at the four $S_4$-invariant momenta and is shown to coincide with the symmetry indicator $z_2$. 
A minimal tight-binding model with $S_4$ symmetry verifies the correspondence between $2P_3$ and $z_2$ throughout its phase diagram. 
The result closes a gap between the topological-field-theory description of the axion response and the topological-band-theory classification by symmetry indicators.
It also extends the known response-indicator equivalence from antiunitary settings such as $C_nT$ symmetry to the unitary, orientation-reversing rotoinversion symmetry $S_4$.
\end{abstract}

\section{\label{sec:introduction}Introduction}
\subsection{\label{subsec:equivalent-ways}Equivalent ways of characterizing time-reversal-symmetric and $C_4T$-symmetric topological insulators}
Topological insulators (TIs) are bulk insulators that support symmetry-protected conducting states on surfaces or edges.
For three-dimensional time-reversal-symmetric band insulators with $\mathcal{T}^2=-1$, the Fu-Kane formula expresses the $\mathbb{Z}_2$ invariant in terms of the time-reversal sewing matrix at the eight time-reversal-invariant momenta (TRIM) $\Gamma_i$ in the Brillouin zone~\cite{fu2006,fu2007a}:
\begin{equation}
(-1)^{\nu_0}=\prod_{i=1}^{8}\frac{\sqrt{\det w(\Gamma_i)}}{\operatorname{Pf}w(\Gamma_i)},
\qquad
w_{mn}(\mathbf{k})\equiv\langle u_m(-\mathbf{k})|\mathcal{T}|u_n(\mathbf{k})\rangle,
\label{eq:FK_formula}
\end{equation}
where $m,n$ label occupied bands, $\operatorname{Pf}$ denotes the Pfaffian, and the square root is chosen continuously throughout the Brillouin zone in a continuous gauge.
The index $\nu_0=1$ identifies a strong topological insulator, whereas $\nu_0=0$ includes both trivial and weak topological insulators.

In the response-theory description introduced by Qi et al.~\cite{qi2008} and developed for crystalline insulators by Essin et al.~\cite{essin2009}, the electromagnetic effective action of a three-dimensional insulator consists of a Maxwell term and a topological term. 
The topological term is proportional to $\theta$, which is quantized to 0 or $\pi$ modulo $2\pi$ under time-reversal (TR) symmetry~\cite{qi2008,essin2009}.
$\theta$ is expressed as a Chern-Simons integral over the Brillouin zone:
\begin{equation}
\theta=2\pi P_3 = \frac{1}{8\pi}\int d^3\bm{k} \epsilon^{ijk}
\text{Tr}\left[\left(\mathcal{F}^{ij}(\bm{k}) - \frac{2}{3}\text{i} \mathcal{A}^i(\bm{k}) \mathcal{A}^j(\bm{k})\right) \mathcal{A}^k(\bm{k})\right]
\label{P3_def}
\end{equation}
where $\mathcal{A}^i_{mn}(\bm{k}) = -\text{i}\langle u_m(\bm{k})|\partial_{k_i}|u_n(\bm{k})\rangle$ is the non-Abelian Berry connection 
and $\mathcal{F}^{ij}(\bm{k}) = \partial_{k_i}\mathcal{A}^j(\bm{k}) - \partial_{k_j}\mathcal{A}^i(\bm{k}) + \text{i}[\mathcal{A}^i(\bm{k}),\mathcal{A}^j(\bm{k})]$ is the non-Abelian Berry curvature. 
The indices $m,n$ run over all occupied bands. 
Therefore, all TR-symmetric insulators fall into two distinct classes, with $\theta=0$ or $\pi$, corresponding to trivial and topological insulators, respectively.

The invariants $\nu_0$ and $\theta$ provide two ways to classify a topological phase: the former is obtained from a discrete formula based on bands at high-symmetry momenta, whereas the latter is obtained from a continuous integral over the Brillouin zone.
It has already been proved that the band invariant $\nu_0$ is equivalent to the Chern-Simons expression $\theta/\pi$, establishing a connection between topological band theory (TBT) and topological field theory (TFT)~\cite{qi2011,wang2010}.
This equivalence bridges the gap between TBT and TFT for TR-symmetric topological insulators.

Even without TR symmetry, $\theta$ may still be quantized by other symmetries such as inversion symmetry~\cite{turner2012,fang2012a}, and insulators with $\theta=0$ are trivial, whereas those with $\theta=\pi$ are termed axion insulators~\cite{varnava2020}.
For $C_4T$-symmetric topological insulators, Li and Sun~\cite{li2020} derived a high-symmetry-momentum formula for $\theta$ that generalizes the Fu-Kane formula in Eq.~\eqref{eq:FK_formula}.
This generalized expression from the TBT perspective is shown to be equivalent to the Chern-Simons integral in Eq.~\eqref{P3_def} from the TFT perspective~\cite{li2020,chen2025}, which bridges the gap between TBT and TFT for $C_4T$-symmetric axion insulators.

\subsection{Symmetry-based indicators for axion insulators}
Beyond formulas tailored to a particular symmetry, such as TR and $C_4T$, systematic methods have been proposed for classifying topology from symmetry representations at high-symmetry momenta. They are known as symmetry-based indicators (SIs) and are briefly reviewed below.

The momentum-space constraints on these representations were analyzed through band-structure combinatorics, establishing a connection to the $K$-theory classification of crystalline insulators with unitary symmetries~\cite{kruthoff2017}.
SIs identify band topology by comparing the symmetry representations of occupied bands with those compatible with atomic insulators~\cite{po2017}, while topological quantum chemistry provides a complementary framework based on band representations induced from localized atomic orbitals~\cite{bradlyn2017}.
These approaches extend symmetry-eigenvalue diagnoses to general space-group settings, with explicit mappings from symmetry data to topological indices and anomalous boundary signatures~\cite{khalaf2018a,song2018}.
For magnetic systems, magnetic topological quantum chemistry extends this framework to magnetic space groups and provides the corresponding symmetry-based indicators~\cite{elcoro2021}.
The relevance of such diagnoses to axion-insulator phenomenology has also been explored in magnetic materials, including antiferromagnetic $\mathrm{MnBi}_6\mathrm{Te}_{10}$~\cite{jo2020}.
Together, these developments establish a systematic framework within TBT for identifying topological phases, including axion phases, from symmetry data at high-symmetry momenta.

\subsection{Axion insulators protected by unitary $S_4$ symmetry}
Beyond the discrete indicators in Sec.~\ref{subsec:equivalent-ways} specifically designed for TR and $C_4T$ symmetries, SIs can identify axion phases protected by other symmetries.
One example is the type-I double magnetic space group 81.33, $P\bar{4}$, generated by four-fold rotoinversion $S_4$, whose SI group is $\mathbb{Z}_4\times\mathbb{Z}_2^2$.
It contains one $\mathbb{Z}_4$ indicator, $z_{4S}$, and two $\mathbb{Z}_2$ indicators, $\delta_{2S}$ and $z_2$, all of which are expressed in terms of the band data at high-symmetry momenta.
Among them, $z_2$ is~\cite{elcoro2021}
\begin{eqnarray}
z_2&&=\sum_{K\in K^4}\frac{n_K^{1/2}-n_K^{-3/2}}{2}\text{ mod }2,\nonumber\\
K^4&&=\{\Gamma, Z, M, A\}
\label{eq:SI}
\end{eqnarray}
where $n_K^{\pm1/2}$ and $n_K^{\pm3/2}$ count occupied states with $S_4$ eigenvalues $e^{\mp i\pi/4}$ and $e^{\mp i3\pi/4}$, respectively, at $K\in K^4$.
For vanishing three-dimensional Chern numbers, $\{z_{4S},\delta_{2S},z_2\}=\{0,0,1\}$ diagnoses an axion insulator with $\theta=\pi$. 
Therefore, $z_2$ is a symmetry-based indicator for axion insulators protected by $S_4$ symmetry, derived from the TBT perspective.

As mentioned in Sec.~\ref{subsec:equivalent-ways}, the indicators for $C_4T$-symmetric axion insulators have been explicitly shown to be equivalent to $\theta/\pi$ as defined by the Chern-Simons integral~\cite{li2020,chen2025}.
However, the relation between $z_2$ and the Chern-Simons axion invariant $\theta/\pi$ has not been derived for axion insulators protected by $S_4$ symmetry.
We emphasize that $S_4$ symmetry is unitary, so the previous proof for TR or $C_4T$ symmetry does not apply without modification.
It is therefore necessary to explicitly establish the equivalence between $z_2$ and $\theta/\pi$ for $S_4$-symmetric axion insulators, which connects the TBT and TFT perspectives.

In this work, the equivalence established for $S_4$ symmetry closes a conceptual gap between two standard languages for the axion phase under unitary symmetries: the TFT of the magnetoelectric response and the diagnosis using symmetry-based indicators derived from TBT.
The methods used in this work can inspire further studies on the equivalence between SIs and axion invariants for other unitary symmetries.
Section~\ref{sec:main} expresses $2P_3=\theta/\pi$ as the winding of the $S_4$ sewing matrix.
Sections~\ref{sec:block_decomposition} and \ref{sec:su2_reduction} reduce the sewing matrix to $U(2)$ and $U(1)$ blocks and then to $SU(2)$ blocks, respectively.
Section~\ref{sec:winding_indicator} reduces the winding to a modulo-two mapping degree and evaluates it by counting $S_4$-invariant momenta with the relevant $S_4$ eigenvalues, which gives Eq.~\eqref{eq:SI}.
Section~\ref{sec:tight_binding} verifies the equivalence in a tight-binding model, and Sec.~\ref{sec:conclusion} summarizes the result.

\section{\label{sec:main}Notation and preliminaries}

We take a cubic lattice with primitive lattice vectors $\bm{a}_1=(1,0,0)$, $\bm{a}_2=(0,1,0)$, and $\bm{a}_3=(0,0,1)$.
The $C_2$-invariant lines, defined modulo reciprocal lattice vectors, are $\Gamma-Z$, $X-R$, $M-A$, and $Y-T$, with high-symmetry points
\[
\begin{array}{cccc}
    \Gamma:(0,0,0),\quad X:(\pi,0,0),\quad Y:(0,\pi,0),\quad M:(\pi,\pi,0)\\
    Z:(0,0,\pi),\quad R:(\pi,0,\pi),\quad T:(0,\pi,\pi),\quad A:(\pi,\pi,\pi)
\end{array}
\]
The four $S_4$-invariant momenta are $K^4=\{\Gamma, M, A, Z\}$.

For an insulator with $N$ occupied bands, the $S_4$ sewing matrix is
\begin{equation}
B_{mn}(\bm{k}) = \langle u_m(S_4\bm{k})|S_4|u_n(\bm{k})\rangle
\label{B_def}
\end{equation}
where $m,n$ are occupied-band indices, $S_4\bm{k}=(k_y,-k_x,-k_z)$, and $|u_n(\bm{k})\rangle$ is the periodic part of the Bloch function for the $n$th occupied band.
The sewing matrix for $C_2\equiv(S_4)^2$ is
\begin{equation}
D_{mn}(\bm{k}) = \langle u_m(C_2\bm{k})|C_2|u_n(\bm{k})\rangle
\label{D_def}
\end{equation}
where $C_2\bm{k}=(-k_x,-k_y,k_z)$. The matrices $B(\bm{k})$ and $D(\bm{k})$ are unitary. 

We assume $(S_4)^4=-1$, appropriate for spin-$1/2$ electrons.
Multiplication of $S_4$ by the phase $e^{i\pi/4}$ maps this convention to $(S_4)^4=1$ and leaves the final results unchanged.

Appendix~\ref{app:sec1} shows that, for an $S_4$-symmetric system,
\begin{equation}
  2P_3 =-\frac{1}{24\pi^2}\int d^3\bm{k} \epsilon^{ijk}\text{Tr}\left[(B^\dagger\partial_iB)(B^\dagger\partial_jB)(B^\dagger\partial_kB)\right]\nonumber\text{ mod }2,
\label{eq:P3_with_B}
\end{equation}
where $P_3=\theta/(2\pi)$ and $\partial_i\equiv\partial/\partial k_i$.
In a smooth periodic gauge, $B(\bm{k})$ defines a map $T^3\to U(N)$, and the normalized integral on the right-hand side is its integer-valued three-dimensional winding number, up to an overall sign~\cite{fang2012a}.
Consequently, $2P_3$ equals the parity of this winding number and takes the values $0$ or $1$ modulo two.

\section{\label{sec:block_decomposition}Block decomposition of the $S_4$ sewing matrix at $z_{4S}=\delta_{2S}=0$}

Within the Chern-trivial axion sector considered here, $z_{4S}=\delta_{2S}=0$.
This section shows that, after adding atomic bands, the sewing matrix can be decomposed into $1\times1$ and $2\times2$ blocks.
Atomic bands are topologically trivial, and scalar $1\times1$ blocks give zero in Eq.~\eqref{eq:P3_with_B}.
Only the $2\times2$ blocks can contribute to $2P_3$. We denote the corresponding $S_4$ and $C_2$ sewing matrices by $B_r(\bm{k})$ and $D_r(\bm{k})$, respectively, and focus on these blocks in the subsequent derivation.
The following discussion uses concepts from topological quantum chemistry introduced in the Supplemental Material of Ref.~\cite{elcoro2021}.
We first show that, after adding atomic bands, a symmetry-data vector $\mathbf{B}$ can be decomposed into a sum of single-band and two-band symmetry-data vectors.
We then use this decomposition to write the $S_4$ sewing matrix $B(\bm{k})$ as a direct sum of single-band and two-band sewing matrices. Throughout, the bold symbol $\mathbf{B}$ denotes a symmetry-data vector, whereas $B(\bm{k})$ denotes a sewing matrix.

\subsection{\label{subsec:symmetry_data_vectors}Preparation for the decomposition}
The space group discussed here is type-I double magnetic space group 81.33 $P\bar{4}$, generated by four-fold rotoinversion $S_4$ and lattice translations. 
We introduce the symmetry-data notation and constraints for this group below.

\paragraph{Little groups and small corepresentations.}
At each momentum, the symmetry properties of Bloch states are described by the little group and its small corepresentations~\cite{elcoro2021}.
The little group at a momentum $\bm k$ consists of the space-group operations that leave $\bm k$ invariant modulo a reciprocal lattice vector.
Its small corepresentations describe how Bloch states at $\bm{k}$ transform under these operations. In the present type-I group, no antiunitary operations are present, so small corepresentations coincide with small representations.
In our case, at each $S_4$-invariant momentum $K\in\{\Gamma,A,M,Z\}$, the four one-dimensional double-valued small corepresentations $\bar K_5,\bar K_6,\bar K_7,\bar K_8$ have $S_4$ eigenvalues $e^{3i\pi/4},e^{-i\pi/4},e^{-3i\pi/4},e^{i\pi/4}$, respectively.
At $R$ and $X$, the little co-group is generated by $C_2=S_4^2$; the labels $\bar R_3,\bar R_4, \bar X_3, \bar X_4$ correspond to $C_2$ eigenvalues $-i,+i,-i,+i$, respectively. We use the labels in the imported MBANDREP table without any row permutation.
The small corepresentations can be obtained from the COREPRESENTATIONS tool of the Bilbao Crystallographic Server, \url{https://cryst.ehu.es/cgi-bin/cryst/programs/corepresentations.pl}. 

\paragraph{Symmetry-data vectors.}
Since we consider spin-$1/2$ electrons, only double-valued small corepresentations occur. 
We therefore write a symmetry-data vector as
\begin{widetext}
\begin{equation}
    \begin{aligned}
\mathbf{B}=\bigl(&m(\bar\Gamma_5),\dots,m(\bar\Gamma_8),m(\bar A_5),\dots,m(\bar A_8),m(\bar M_5),\dots,m(\bar M_8),m(\bar Z_5),\dots,m(\bar Z_8),\\
&m(\bar R_3),m(\bar R_4),m(\bar X_3),m(\bar X_4)\bigr)^{\!\top}\in\mathbb{Z}^{20}_{\ge 0},
    \end{aligned}
\end{equation}
\end{widetext}
where each entry is the multiplicity of the corresponding small corepresentation at a high-symmetry momentum. 
Here $\mathbb{Z}^{20}_{\ge 0}$ is the set of 20-component vectors with nonnegative integer entries.
These vectors record the symmetry content of a set of bands at the listed high-symmetry momenta.
For example, a single band with eigenvalues
\[
S_4=e^{3i\pi/4} \text{ at } \Gamma, M\quad S_4=e^{-i\pi/4} \text{ at } A, Z \quad C_2=-i \text{ at } R, X
\] 
has 1 in the entries corresponding to $\bar\Gamma_5$, $\bar A_6$, $\bar M_5$, $\bar Z_6$, $\bar R_3$, $\bar X_3$, and 0 in all other entries.

\paragraph{Compatibility relations.}
Physical band structures must satisfy the compatibility relations. 
For instance, $\Gamma-Z$ is a $C_2$-invariant line, and the multiplicity of each $C_2$ eigensector is constant along the line for an isolated band subspace. 
In other words, the bands at $\Gamma$ and $Z$ must satisfy
\[
\begin{aligned}
m(\bar\Gamma_5)+m(\bar\Gamma_6)&=m(\bar Z_5)+m(\bar Z_6),\\
m(\bar\Gamma_7)+m(\bar\Gamma_8)&=m(\bar Z_7)+m(\bar Z_8).
\end{aligned}
\]
These equations match the multiplicities of the two $C_2=(S_4)^2$ eigensectors at $\Gamma$ and $Z$.
The eight compatibility relations are encoded by an integer matrix $CR\in\mathbb{Z}^{8\times 20}$:
\begin{equation}
CR\cdot\mathbf{B}=0.
\label{eq:CR_condition}
\end{equation}

\paragraph{Symmetry-based indicators.}
The two symmetry-based indicators $z_{4S}$ and $\delta_{2S}$ are calculated from the $S_4$ or $C_2$ eigenvalues at high-symmetry momenta~\cite{elcoro2021}:
\begin{equation}
\begin{aligned}
z_{4S}&=-\frac12 n_Z^{1/2}+\frac12 n_Z^{-1/2}-\frac32n_Z^{3/2}+\frac32n_Z^{-3/2}\\
&\quad-\frac12 n_A^{1/2}+\frac12 n_A^{-1/2}-\frac32n_A^{3/2}+\frac32n_A^{-3/2}+n_R^{1/2}-n_R^{-1/2}\text{ mod }4,\\
\delta_{2S}&=-n_Z^{3/2}+n_Z^{-3/2}-n_A^{3/2}+n_A^{-3/2}+n_\Gamma^{3/2}-n_\Gamma^{-3/2}+n_M^{3/2}-n_M^{-3/2}\text{ mod }2
\end{aligned}
\end{equation}
where $n_{K}^{\frac12,-\frac12,\frac32,-\frac32}$ count occupied states at $K$ with $S_4$ eigenvalues $e^{-i\frac{\pi}{4}},e^{i\frac{\pi}{4}},e^{-i\frac{3\pi}{4}},e^{i\frac{3\pi}{4}}$, respectively, and $n_{R}^{\frac12,-\frac12}$ count occupied states at $R$ with $C_{2}$ eigenvalues $e^{-i\frac{\pi}{2}},e^{i\frac{\pi}{2}}$, respectively.
The eigenvalue multiplicities are entries of the symmetry-data vector, so these indicators are linear forms on $\mathbf{B}$ evaluated modulo their respective orders. In this convention, $n_R^{1/2}=m(\bar R_3)$ and $n_R^{-1/2}=m(\bar R_4)$, so the $R$-entries of $\mathbf p_1$ are $(+1,-1)$ and those of $P_1=2\mathbf p_1$ are $(+2,-2)$.
Writing $\mathbf p_1,\mathbf p_2$ for the coefficient vectors in these formulas, with half-integer entries in $\mathbf p_1$ and integer entries in $\mathbf p_2$, we have
\begin{equation}
z_{4S}(\mathbf B)\equiv \mathbf p_1\!\cdot\!\mathbf B \pmod 4,\qquad
\delta_{2S}(\mathbf B)\equiv \mathbf p_2\!\cdot\!\mathbf B \pmod 2 .
\end{equation}
Clearing denominators with the integral rows $P_1=2\mathbf p_1$ and $P_2=\mathbf p_2$, the conditions $z_{4S}=\delta_{2S}=0$ become the congruences
\begin{equation}
P_1\cdot\mathbf B\equiv 0 \pmod 8,\quad
P_2\cdot\mathbf B\equiv 0 \pmod 2. 
\label{eq:congruence_condition}
\end{equation}

\paragraph{The symmetry-data monoid.}
We collect all admissible data into the set
\begin{equation}
\mathcal{M}=\bigl\{\mathbf B\in\mathbb{Z}^{20}_{\ge 0}\Big|CR\cdot\mathbf B=0,P_1\cdot\mathbf B\equiv 0\text{ mod}(8),P_2\cdot\mathbf B\equiv 0\text{ mod}(2)\bigr\}.
\label{eq:M_def}
\end{equation}
Because $\mathcal{M}$ is closed under addition and contains $0$, it is a commutative monoid: the symmetry-data monoid of $S_4$-symmetric systems with $z_{4S}=\delta_{2S}=0$. 

\paragraph{Elementary band corepresentations (EBRs).}
A band corepresentation describes the symmetry transformation of Bloch states constructed from localized orbitals placed at symmetry-related positions in a crystal, forming a Wyckoff orbit.
Physically, these states can be viewed as bands originating from an array of atomic orbitals, whose local symmetry properties determine the resulting band corepresentation~\cite{elcoro2021}.
An elementary band corepresentation (EBR) is an indivisible building block of such atomic band corepresentations: it cannot be decomposed into a direct sum of smaller band corepresentations.
Taken as a complete set of occupied bands, an EBR admits symmetry-respecting, exponentially localized Wannier functions and represents an atomic insulator, with vanishing symmetry-based indicators and axion invariant.
Let $\mathcal{E}\in\mathbb{Z}^{20\times 18}_{\ge 0}$ denote the EBR matrix for this group, obtained from MBANDREP (\url{https://cryst.ehu.es/cgi-bin/cryst/programs/mbandrep.pl}). Each column $\mathcal{E}_{j}$ records the symmetry data of an atomic band representation and belongs to $\mathcal{M}$, because EBRs satisfy compatibility relations and have vanishing indicators.
The source table contains 36 columns, of which 18 correspond to double-valued representations; only these columns are retained in $\mathcal E$.

The decomposition proof proceeds as follows.
We enumerate all single-band and two-band symmetry-data vectors in $\mathcal{M}$, obtaining $B_1^{(1)},\dots,B_1^{(16)}$ and $B_2^{(1)},\dots,B_2^{(438)}$, respectively.
We collect these vectors as the columns of $G\in\mathbb{Z}^{20\times 454}_{\ge0}$:
\begin{equation}
G\equiv[B_1^{(1)},\cdots,B_1^{(16)},B_2^{(1)},\cdots,B_2^{(438)}].
\end{equation} 
We then establish that, for every $\mathbf B\in\mathcal{M}$, there exist nonnegative integer vectors $\mathbf w\in\mathbb{Z}^{454}_{\ge0}$ and $\mathbf t\in\mathbb{Z}^{18}_{\ge0}$ such that
\begin{equation}
    \mathbf B + \mathcal{E}\mathbf t = G\mathbf w.
    \label{eq:B_decomposition}
\end{equation}
Thus, after adding atomic bands, any symmetry-data vector in $\mathcal M$ can be written as a sum of single-band and two-band symmetry-data vectors.

\subsection{\label{subsec:one_two_band_reduction}Reduction to symmetry data of single-band and two-band systems}

Gordan's lemma reduces the infinite family of decomposition problems in Eq.~\eqref{eq:B_decomposition} to a finite check~\cite{Bruns2009}.
To apply it, we rewrite Eq.~\eqref{eq:M_def} as
\begin{equation}
    \mathcal{M}=\bigl\{\mathbf{x}\in\mathbb{R}^{20}_{\ge0}\Big|CR\cdot\mathbf x=0\bigr\}\cap\bigl\{\mathbf x'\in\mathbb{Z}^{20}\Big|P_1\mathbf x'\equiv0\pmod8,\quad P_2\mathbf x'\equiv0\pmod2 \bigr\}.
\end{equation}
The monoid $\mathcal M$ is the intersection of a pointed rational polyhedral cone with a lattice defined by the two congruence conditions. 
Gordan's lemma states that this monoid is finitely generated: every element is a nonnegative integer combination of a finite set of generators.

Using Normaliz~\cite{bruns2010}, we obtain the Hilbert basis
\begin{equation}
H=\{h_1,\dots,h_s\},\qquad s=2466 .
\end{equation}
An arbitrary $\mathbf B\in\mathcal M$ can therefore be written as
\begin{equation}
\mathbf B=\sum_{i=1}^{s} n_i\, h_i,\qquad n_i\in\mathbb{Z}_{\ge0}. 
\label{eq:B_basis_expansion}
\end{equation}

Concretely, for each $i$ we solve the integer feasibility problem
\begin{equation}
Gw_i - \mathcal{E}t_i = h_i,\quad w_i\in\mathbb{Z}^{454}_{\ge0},\ t_i\in\mathbb{Z}^{18}_{\ge0}.
\label{eq:feasibility_problem}
\end{equation}
All $s=2466$ systems in Eq.~\eqref{eq:feasibility_problem} admit nonnegative integer solutions, which we verify by exact substitution. For the expansion in Eq.~\eqref{eq:B_basis_expansion}, set
\begin{equation}
\mathbf w=\sum_i n_i w_i\ \ge 0,\qquad \mathbf t=\sum_i n_i t_i\ \ge 0 .
\end{equation}
Then
\begin{equation}
G\mathbf w=\Bigl(\sum_i n_i h_i\Bigr)+\mathcal{E}\Bigl(\sum_i n_i t_i\Bigr)=\mathbf B+\mathcal{E}\,\mathbf t ,
\end{equation}
which is precisely Eq.~\eqref{eq:B_decomposition}.

\subsection{\label{subsec:sewing_block_decomposition}Block decomposition of the $S_4$ sewing matrix}

The preceding subsection shows that, after adding EBRs, any symmetry-data vector in $\mathcal M$ becomes a sum of single-band and two-band symmetry-data vectors.
The corresponding band groups each satisfy $(z_{4S},\delta_{2S})=(0,0)$.
This gives a decomposition of the $S_4$ sewing matrix $B(\bm k)$ into $1\times1$ and $2\times2$ blocks after adding atomic bands.

These added atomic bands have symmetry data $\mathcal E\mathbf t$ and do not contribute to $2P_3$. The stabilized sewing matrix can therefore be written as
\begin{equation}
B(\mathbf k)\ \simeq\ \bigoplus_{i}\underbrace{B^{(1)}_i(\mathbf k)}_{1\times 1}\ \oplus\ \bigoplus_{j}\underbrace{B^{(2)}_j(\mathbf k)}_{2\times 2}, 
\end{equation}
in which every $1\times1$ block and every $2\times2$ block individually satisfies $z_{4S}=\delta_{2S}=0$, and any nontrivial $z_2$ is carried by the $2\times2$ blocks alone. 
The contribution of a $1\times1$ scalar block to Eq.~\eqref{eq:P3_with_B} vanishes.
Equation~\eqref{eq:SI} also gives $z_2=0$ for all 16 single-band symmetry-data vectors, so $1\times1$ blocks do not contribute to $z_2$. 
Hence, no block larger than $2\times2$ is required for the symmetry-data decomposition, and only the $2\times2$ blocks are relevant to $P_3$.
In the following sections, we denote the $2\times2$ sewing-matrix blocks by $B_r(\bm k)$.

\section{\label{sec:su2_reduction}Reducing $B_r(\bm k)$ from $U(2)$ to $SU(2)$}

As shown in Sec.~\ref{sec:block_decomposition}, the nontrivial winding relevant to $2P_3$ is carried by two-band blocks $B_r(\bm{k})\in U(2)$.
To establish the equivalence between $2P_3$ and the indicator $z_2$, we next reduce these $U(2)$ blocks to $SU(2)$ blocks.

For a generic unitary $S_4$ block, $\det B_r(K)$ at an $S_4$-invariant momentum is gauge invariant and cannot be changed by a smooth gauge transformation.
For example, $\det B_r(\Gamma)=1$ and $\det B_r(A)=-1$ are gauge invariant, and no gauge transformation can give an $SU(2)$ sewing matrix.
After adding atomic bands, however, all relevant two-band data can be rewritten in terms of determinant-one two-band generators.
For these generators the sewing matrices are represented by $SU(2)$-valued blocks.

Below we first give the criterion for an individual two-band block to be reduced directly to an $SU(2)$-valued sewing matrix.
Two-band blocks that fail this criterion can nevertheless be handled after adding atomic bands, because the symmetry data can be reblocked into determinant-one two-band generators.

\subsection{\label{subsec:direct_su2_reduction}Direct reduction of $B_r(\bm{k})$ to $SU(2)$}

\subsubsection{Criteria for direct reduction from $U(2)$ to $SU(2)$}

Gauge invariance imposes necessary conditions for reducing $B_r(\bm k)$ directly to $SU(2)$ by a gauge transformation.

Under the standing assumption of vanishing 3D Chern numbers, the occupied bands may be chosen in a smooth periodic gauge.
Let $U(\bm{k})=(|u_1(\bm{k})\rangle,|u_2(\bm{k})\rangle)$ be a smooth frame for the occupied bands.
For a $U(2)$ gauge transformation $G$, the occupied states transform as $U(\bm{k})\rightarrow U(\bm{k})G(\bm{k})$.
Then the sewing matrix transforms as $B_r'(\bm{k}) = G^\dagger(S_4\bm{k})B_r(\bm{k})G(\bm{k})$.
Writing $\det G(\bm{k})=e^{i\alpha(\bm{k})}$, we obtain $\det B_r'(\bm{k})=1$ if
\begin{equation}
i(\alpha(S_4\bm{k})-\alpha(\bm{k}))=\ln\det B_r(\bm{k})
\label{eq:alpha_condition}
\end{equation}
At an $S_4$-invariant momentum $K\in\{\Gamma,Z,M,A\}$, this transformation leaves the determinant unchanged.
Thus, $\det B_r(K)=1$ at every such $K$ is a necessary condition for direct reduction to $SU(2)$.

Furthermore, this transformation is constrained at $X(\pi,0,0)$ and $Y(0,\pi,0)$ because these two momenta are mapped into each other under $S_4$:
\begin{eqnarray}
i(\alpha(Y)-\alpha(X))&&=\ln\det B_r(X)\\
i(\alpha(X)-\alpha(Y))&&=\ln\det B_r(Y).
\end{eqnarray}
Hence the product $\det B_r(X)\det B_r(Y)$ is also gauge invariant.
From Eq.~\eqref{B_relation} and $(S_4)^4=(C_2)^2=-1$, the matrices $B_r$ and $D_r$ satisfy the exact identities
\begin{equation}
    D_r(\bm{k})=B_r(S_4\bm{k})B_r(\bm{k}),\qquad D_r(C_2\bm{k})D_r(\bm{k})=-I.
    \label{eq:BD_identities}
\end{equation}
This gives the gauge-invariant quantity $\det D_r(X)=\det B_r(Y)\det B_r(X)=\det D_r(Y)$.
If $B_r$ is $SU(2)$-valued, this product equals 1. Thus, $\det D_r(X)=1$ is another necessary condition.

The necessary conditions are therefore
\begin{equation}
\det B_r(K)=1,\qquad \det D_r(X)=1
\label{eq:reducible_condition}
\end{equation}
The determinant is constant along each $C_2$-invariant line. Since $S_4$ relates $X$ and $Y$, the condition $\det D_r(X)=1$ also implies $\det D_r(Y)=\det D_r(R)=\det D_r(T)=1$.
The conditions in Eq.~\eqref{eq:reducible_condition} are also sufficient to solve Eq.~\eqref{eq:alpha_condition}, as argued below.

\subsubsection{Constructing a gauge transformation from $U(2)$ to $SU(2)$}

We seek a gauge transformation $G(\bm{k})$, with $\det G(\bm{k})=e^{i\alpha(\bm{k})}$, that satisfies Eq.~\eqref{eq:alpha_condition}.
The construction first removes the winding of the determinant and then its remaining phase, using Eq.~\eqref{eq:reducible_condition}.

Let $\bm{m}=(m_x,m_y,m_z)\in\mathbb{Z}^3$ denote the winding numbers of $\det B_r(\bm{k})$ along the three fundamental loops of the Brillouin torus, and let $\bm{n}=(n_x,n_y,n_z)$ denote those of $\det G(\bm{k})=e^{i\alpha(\bm{k})}$.
The loop $\gamma_z: (0,0,-\pi)\to(0,0,\pi)$ is a $C_2$-invariant line, so the $C_2$ eigenvalue multiplicities and $\det D(\bm{k})$ are constant along it.
Therefore, along $\gamma_z$,
\begin{equation}
\det D(\bm{k})=\det B(-\bm{k})\det B(\bm{k})=\det B(\Gamma)\det B(\Gamma)=1,
\end{equation}
and the winding number of $\det B_r$ satisfies
\begin{equation}
    \begin{aligned}
        m_z&=\frac{1}{2\pi i}\int_{\gamma_z}d\ln\det B_r(\bm{k})\\
        &=\frac{1}{\pi i}\int_{\Gamma}^{Z}d\ln\det B_r(\bm{k})\\
        \Rightarrow(-1)^{m_z}&=\frac{\det B_r(Z)}{\det B_r(\Gamma)}
    \end{aligned}
\end{equation}
Next, consider the loop $\gamma_x:(-\pi,0,0)\to(\pi,0,0)$.
Using Eq.~\eqref{eq:BD_identities}, the winding number of $\det D_r(\bm{k})$ along this loop satisfies
\begin{equation}
\begin{aligned}
-m_y+m_x&=\frac{1}{2\pi i}\int_{\gamma_x}d\ln\det D_r(\bm{k})\\
&=\frac{1}{\pi i}\int_{\Gamma}^{X}d\ln\det D_r(\bm{k})\\
\Rightarrow(-1)^{-m_y+m_x}&=\frac{\det D(X)}{\det D(\Gamma)}
\end{aligned}
\end{equation}
Similarly, along $\gamma_y:(0,-\pi,0)\to(0,\pi,0)$, we obtain $(-1)^{m_y+m_x}=\frac{\det D(Y)}{\det D(\Gamma)}$.
Substituting $\det B(\Gamma)=\det B(Z)=\det D(X)=\det D(Y)=1$ gives
\begin{equation}
m_z,\quad m_x-m_y,\quad m_x+m_y\in 2\mathbb{Z}
\label{eq:B_even_winding}
\end{equation}

Since $S_4(k_x,k_y,k_z)=(k_y,-k_x,-k_z)$, the phase difference on the left-hand side of Eq.~\eqref{eq:alpha_condition} has winding vector $(-n_y-n_x,n_x-n_y,-2n_z)$.
The parity conditions in Eq.~\eqref{eq:B_even_winding} ensure that we can choose a gauge transformation $G_1(\bm{k})$ whose determinant has integer winding numbers
\begin{equation}
n_x=\frac{-m_x+m_y}{2},n_y=\frac{-m_x-m_y}{2},n_z=-\frac{m_z}{2}
\end{equation}
The determinant of the transformed block $B_r'(\bm{k})=G_1^\dagger(S_4\bm{k})B_r(\bm{k})G_1(\bm{k})$ then has zero winding along all three fundamental loops and still satisfies $\det B_r'(K)=1$ at the $S_4$-invariant momenta.
A second gauge transformation $G_2(\bm{k})$ can then remove the remaining phase of $\det B_r'(\bm{k})$.
Therefore, $B_r(\bm{k})$ is reduced to an $SU(2)$-valued sewing matrix by $G(\bm{k})=G_1(\bm{k})G_2(\bm{k})$, provided Eq.~\eqref{eq:reducible_condition} holds.

\subsection{\label{subsec:stable_su2_reduction}Reduction of the remaining blocks after adding atomic bands}

We now explain how to treat two-band blocks that violate Eq.~\eqref{eq:reducible_condition}. 
The obstruction is not a gauge artifact: at an $S_4$-fixed momentum $K$, the sewing matrix changes by conjugation,
\begin{equation}
B_r'(K)=G^\dagger(K)B_r(K)G(K),
\end{equation}
so $\det B_r(K)$ is invariant under any smooth gauge transformation. 
Similarly, at the $C_2$-fixed points $R$ and $X$, the determinant of the corresponding two-band $D_r$ block is fixed by the $C_2$ eigenvalue content. 
Thus, a two-band block from Sec.~\ref{sec:block_decomposition} need not admit a direct reduction to $SU(2)$.

We address this obstruction through stable equivalence, i.e., by adding atomic bands.
Adding an atomic band representation changes neither the Chern-Simons invariant nor the symmetry indicators. 
At the level of symmetry data, this means that a two-band symmetry-data vector $\mathbf b$ can be replaced by
\begin{equation}
\mathbf b+\mathcal E\mathbf v ,
\end{equation}
where $\mathcal E$ is the EBR matrix introduced in Sec.~\ref{sec:block_decomposition} and $\mathbf v\in\mathbb Z_{\ge0}^{18}$ records how many copies of each atomic band representation are added.
The question is then whether this stabilized vector can be decomposed into two-band vectors satisfying Eq.~\eqref{eq:reducible_condition}.
If such a decomposition exists, the original $U(2)$ block together with the added atomic bands can be regrouped into $2\times2$ blocks, each satisfying Eq.~\eqref{eq:reducible_condition}.
Each resulting block then admits the direct reduction to $SU(2)$ described above.

In the convention of Sec.~\ref{sec:block_decomposition}, the symmetry-data vector of a two-band block records its fixed-point multiplicities.
At each $S_4$-invariant momentum $K=\Gamma,A,M,Z$, the conjugate pairs are
\begin{equation}
(\bar K_5,\bar K_7),\qquad (\bar K_6,\bar K_8),
\end{equation}
and $\det B_r(K)=1$ is equivalent to equal multiplicities within each pair. 
At the $C_2$-invariant momenta $R$ and $X$, $\det D_r=1$ is equivalent to equal multiplicities of the conjugate pair
\begin{equation}
(\bar R_3,\bar R_4),\qquad (\bar X_3,\bar X_4).
\end{equation}
Enumeration gives 16 two-band symmetry-data vectors satisfying Eq.~\eqref{eq:reducible_condition}, whose corresponding blocks admit direct reduction to $SU(2)$.
We refer to these as determinant-one two-band generators.
Let $S_2^{(1)},\dots,S_2^{(16)}$ denote those two-band elements of $\mathcal M$, collected as columns of
\begin{equation}
S=\bigl(S_2^{(1)},\dots,S_2^{(16)}\bigr)\in\mathbb Z_{\ge0}^{20\times16}.
\label{eq:det1-two-band-vector}
\end{equation}
The stabilized reduction needed for the proof is the nonnegative integer problem
\begin{equation}
S\mathbf u=\mathbf b+\mathcal E\mathbf v,\qquad
\mathbf u\in\mathbb Z_{\ge0}^{16},\quad \mathbf v\in\mathbb Z_{\ge0}^{18}.
\label{eq:stable_su2_reblocking}
\end{equation}
Equation~\eqref{eq:stable_su2_reblocking} expresses the stabilized two-band data as a sum of determinant-one two-band generators.
Only these determinant-one blocks are used in the $SU(2)$ degree argument in Sec.~\ref{sec:winding_indicator}.

We verified Eq.~\eqref{eq:stable_su2_reblocking} for all 438 admissible two-band symmetry-data vectors $\mathbf b\in\mathcal M$.
Of these, 16 satisfy the determinant-one criterion directly and form the columns of $S$.
Solving Eq.~\eqref{eq:stable_su2_reblocking} for all 438 targets gives a nonnegative integer solution in every case.
Thus, after adding EBRs, the remaining $U(2)$ blocks can also be regrouped into two-band blocks that admit direct reduction to $SU(2)$.

An explicit example is given in Appendix~\ref{app:SU2_example}.

\section{\label{sec:winding_indicator}From the winding of $B$ to the $z_2$ indicator}

\subsection{From winding number to mapping degree after reduction to $SU(2)$ blocks}
As shown in Secs.~\ref{sec:block_decomposition} and \ref{sec:su2_reduction}, after adding trivial atomic bands, the sewing matrix $B(\bm k)\in U(N)$ can be decomposed into $SU(2)$ blocks and $1\times1$ blocks.
Only the $SU(2)$ blocks contribute to the topological invariant or the $z_2$ indicator.
Below, $B_r(\bm{k})$ denotes such a determinant-one block.
Since $SU(2)\simeq S^3$, the winding number in Eq.~\eqref{eq:P3_with_B} is the degree of a map $T^3\to S^3$ for each block.
Thus,
\begin{equation}
    \begin{aligned}
    2P_3 &= -\frac{1}{24\pi^2}\sum_{r}\int d^3\bm{k} \epsilon^{ijk} \text{Tr}[(B_r^\dagger\partial_iB_r)(B_r^\dagger\partial_jB_r)(B_r^\dagger\partial_kB_r)]\text{ mod }2\\
            &=\sum_r\text{deg}_2[B_r]
    \end{aligned}
\label{eq:P3_to_degree}
\end{equation}
where $\text{deg}_2$ denotes the mapping degree modulo 2.
The mapping degree is defined for maps between closed, oriented manifolds of the same dimension, such as $T^3\to S^3$.
In addition to its integral expression, the mapping degree can be obtained by counting the preimages of a regular value of $B_r$, weighted by the sign of the Jacobian at each preimage~\cite{dubrovin1985a,wang2010,li2020}.
We illustrate this idea below using a one-dimensional map. A rigorous mathematical derivation is given in Ref.~\cite{dubrovin1985a}.

\begin{figure}[htbp]
\centering
\includegraphics[width=0.8\textwidth]{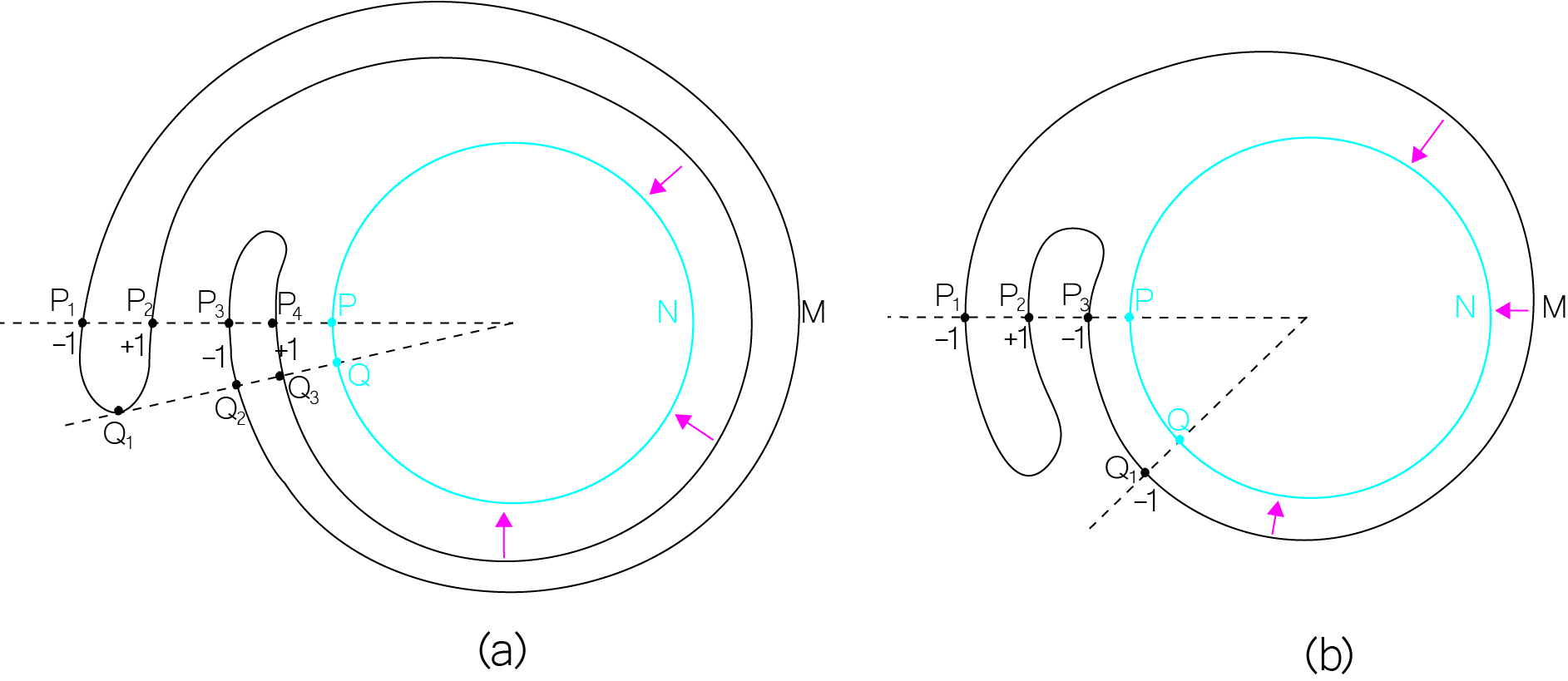}
\caption{Map between two circles $M$ and $N$. The mapping degree is 0 in (a) and 1 in (b). 
The number $\pm1$ beside each preimage point is the sign of the Jacobian at that point.
The counting must be performed at a regular value; point $Q$ in (a) is not regular.
Adapted from Ref.~\cite{wang2010}.}
\label{fig:winding}
\end{figure}

Figure~\ref{fig:winding} illustrates this construction for a map $f:M\to N$ between two copies of $S^1$.
If $\theta$ and $\phi$ are coordinates on $M$ and $N$, respectively, then $\text{deg}(f)=\frac{1}{2\pi}\int_{0}^{2\pi}d\theta\,d\phi/d\theta=n$.
Equivalently,
$\text{deg}(f)=\sum_{f(\theta_i)=\phi_0}\text{sgn}\det(\frac{\partial \phi}{\partial\theta})\Big|_{\theta_i}$.
The value $\phi_0$ must be regular, i.e., the Jacobian at each preimage must be nonsingular.
Point $Q$ in Fig.~\ref{fig:winding}(a) is not regular and gives the incorrect count $\text{deg}(f)=1$, whereas any regular value gives the same degree.

\subsection{Calculating the mapping degree of $B_r(\bm k)$}
In our case, the preimages are Brillouin-zone momenta $\bm{k}\in T^3$, and their images are $B_r(\bm k)\in SU(2)\simeq S^3$.
We evaluate $\text{deg}_2[B_r]$ in Eq.~\eqref{eq:P3_to_degree} using the counting method introduced above:
\begin{equation}
\text{deg}_2[B_r]=\sum_{B_r(\bm k_i)=A}\text{sgn}\det\left(\partial_{\bm k} B_r\right)\Big|_{\bm{k}_i}\pmod2=\sum_{B_r(\bm k_i)=A}1
\end{equation}
where $A$ is an arbitrary $SU(2)$ matrix satisfying $\det\left(\partial_{\bm k} B_r\right)\neq0$. 
In other words, we count the preimages $\bm k$ such that $B_r(\bm k)=A$.
For any regular value $A$, this count equals the mapping degree modulo 2.

For convenience, preimages are counted at some $A=B_r(K)$, where $K$ is an $S_4$-invariant momentum.
As shown in Appendix~\ref{app:sec4}, a gauge transformation can be chosen such that $D_r(\bm{k})$ is the constant matrix $i\sigma_z$.
The sewing matrices satisfy
\begin{equation}
    B_r(S_4\bm{k}) = D_r(\bm{k})B_r^\dagger(\bm{k}).
    \label{eq:B_D_relation_general}
\end{equation}
Hence at $S_4$-invariant momenta,
\begin{equation}
    B_r(K) = D_r(K)B_r^\dagger(K).
    \label{B_D_relation}
\end{equation}
Since $D_r(K)=i\sigma_z$, Eq.~\eqref{B_D_relation} gives
\begin{equation}
    B_r(K) = \pm\frac{1}{\sqrt{2}}(\sigma_0+\text{i}\sigma_z)\equiv A_{\pm}
    \label{eq:A_pm_def}
\end{equation}

Appendix~\ref{app:sec5} shows that $A_\pm$ are not forced by symmetry to be singular values.
If an accidental singularity occurs, a perturbation makes $A_\pm$ regular values without changing the topological invariant.
The modulo-two degree can therefore be evaluated at $A_+$. 
We therefore count the momenta $\bm k$ such that $B_r(\bm{k})=A_+$; this number modulo 2 is $\text{deg}_2[B_r]$.

First, if $\bm{k}$ is not an $S_4$-invariant momentum and $B_r(\bm{k})=A_+$, Eq.~\eqref{eq:B_D_relation_general} gives
\begin{equation}
    B_r(S_4\bm{k})=\text{i}\sigma_z A_+^\dagger=A_+
\end{equation}
Thus the preimages away from the $S_4$-invariant momenta occur in $S_4$-related pairs and do not contribute modulo two.
The same argument applies to preimages of $A_-$.

As a result, we only need to consider $\bm k\in K^4=\{\Gamma,Z,M,A\}$:
\begin{equation}
\text{deg}_2[B_r]=n_r\text{ mod } 2
\label{eq:deg2_count}
\end{equation}
where $n_r$ is the number of $A_+$ values among the four $S_4$-invariant momenta $K\in\{\Gamma, Z, M, A\}$.
Since $D_r=i\sigma_z$ constrains $B_r(K)$ to be diagonal, its diagonal elements are the $S_4$ eigenvalues of the two bands.
The condition $B_r(K)=A_+$ therefore corresponds to the pair of $S_4$ eigenvalues $e^{\pm i\pi/4}$, whereas $B_r(K)=A_-$ corresponds to $e^{\pm i3\pi/4}$.
Thus, a block with $B_r(K)=A_+$ contributes one to $n_K^{1/2}$ in the $z_2$ indicator, Eq.~\eqref{eq:SI}, which counts the bands with $S_4$ eigenvalue $e^{-i\frac{\pi}{4}}$ at $K$.
Similarly, a block with $B_r(K)=A_-$ contributes one to $n_K^{-3/2}$. Since symmetry constrains $B_r(K)$ to be $A_\pm$, each block takes the value $A_-$ at the remaining $4-n_r$ invariant momenta.
Combining all $2\times2$ blocks gives
\begin{equation}
    \begin{aligned}
    z_2&=\sum_K\frac{n_K^{1/2}-n_K^{-3/2}}{2}\text{ mod }2\\
    &=\sum_r\frac{n_r-(4-n_r)}{2}\text{ mod }2\\
    &=\sum_r\text{deg}_2[B_r]\text{ mod }2=2P_3.
    \end{aligned}
\end{equation}
Thus, the equivalence is proven.

\section{\label{sec:tight_binding}Tight-binding model}

Below we construct a tight-binding model that realizes the $S_4$-protected axion phase and numerically verify the equivalence between $2P_3$ and $z_2$. 
Varying the model parameters yields a topological phase diagram that provides a more complete physical picture.

To match the setting of the $z_2$ symmetry indicator~\cite{elcoro2021}, the model belongs to space group $P\bar 4$, with $S_4$ as the only symmetry that quantizes $\theta$.
Other symmetries, such as inversion $\mathcal I$, would independently quantize $\theta$ to $0$ or $\pi$ if present~\cite{turner2012,fang2012a}, concealing the $S_4$-specific contribution to the axion response.
The model here is adapted from Ref.~\cite{tanaka2022} and is defined on a simple tetragonal lattice.
In the basis $(c_{A\uparrow},c_{A\downarrow},c_{B\uparrow},c_{B\downarrow})$, with $\tau$ acting on the orbital and $\sigma$ on the spin index, the Hamiltonian is
\begin{eqnarray}
H(\bm{k})&&=\epsilon(\bm{k})\,\tau_x\otimes\sigma_0 + u\sin k_x\,\tau_y\otimes\sigma_x + u\sin k_y\,\tau_y\otimes\sigma_y \nonumber\\
&&\quad+ v\sin k_z\,\tau_y\otimes\sigma_z + g(\bm{k})\,\tau_z\otimes\sigma_0 + m_z\,\tau_0\otimes\sigma_z, \nonumber\\
\epsilon(\bm{k})&&=\epsilon_0-\cos k_x-\cos k_y+2\cos k_z,\nonumber\\
g(\bm{k})&&=\eta\,(\cos k_x-\cos k_y)+\kappa\sin k_z,\nonumber\\
U_{S_4}&&=\tau_x\otimes e^{-\text{i}\frac{\pi}{4}\sigma_z},\qquad S_4\bm{k}=(k_y,-k_x,-k_z).
\label{eq:TB_model}
\end{eqnarray}
$U_{S_4}$ is the representation of $S_4$ in the orbital basis and satisfies $U_{S_4}^4=-1$.

The first four terms form an $S_4$-symmetric Dirac Hamiltonian.
The matrix $\tau_z$ is odd under $U_{S_4}$ and the form factor $g(\bm{k})$ is odd under $\bm{k}\to S_4\bm{k}$, so $g\tau_z$ is $S_4$ symmetric. 
The Zeeman term $m_z\sigma_z$ is also $S_4$-symmetric, breaking time-reversal symmetry and opening a gap at the Dirac points. 
The enumeration of cubic-lattice point symmetries in Appendix~\ref{app:TB_symmetry_audit} includes more candidate symmetries than those of the tetragonal space group $P\bar{4}$ and shows that the model has no additional symmetries beyond those generated by $S_4$, including $C_2\equiv(S_4)^2$.

Throughout this section and Appendix~\ref{app:TB_details} we use $\epsilon_0=2.4$, $m_z=0.3$, $\eta=\kappa=0.5$, and $u=v=1$ unless stated otherwise, and fill the two lower bands.
As a result, $u=v$ can be used as the energy unit.

In Appendix~\ref{subsec:symmetry_data} we compute the symmetry indicators of the toy model via Eq.~\eqref{eq:SI}, as well as $2P_3$ by counting the preimages of $B(\mathbf{k})=A_+$ at the four $S_4$-invariant momenta, as indicated in Eqs.~\eqref{eq:P3_to_degree} and \eqref{eq:deg2_count}.
The sewing matrices at $K\in\{\Gamma, M, A, Z\}$ take the form of Eq.~\eqref{eq:A_pm_def}. Both the symmetry indicator $z_2$ and $2P_3$ obtained by preimage counting equal $1$, confirming the axion phase of the model for the parameters above.

The spectrum, symmetry analysis, and numerical methods are detailed in Appendix~\ref{app:TB_details}.
Furthermore, by analyzing the symmetry-protected Weyl points of this model, Appendix~\ref{subsec:weyl_points} indicates that the two occupied bands can never be decomposed into two single bands.
This provides a physical interpretation of why the topological axion-insulator phase requires contributions from irreducible $SU(2)$ band blocks, whereas $1\times1$ blocks do not contribute.
This is consistent with the mathematical analysis in Sec.~\ref{sec:block_decomposition}.

\begin{figure}[htbp]
    \centering
    \includegraphics[width=\textwidth]{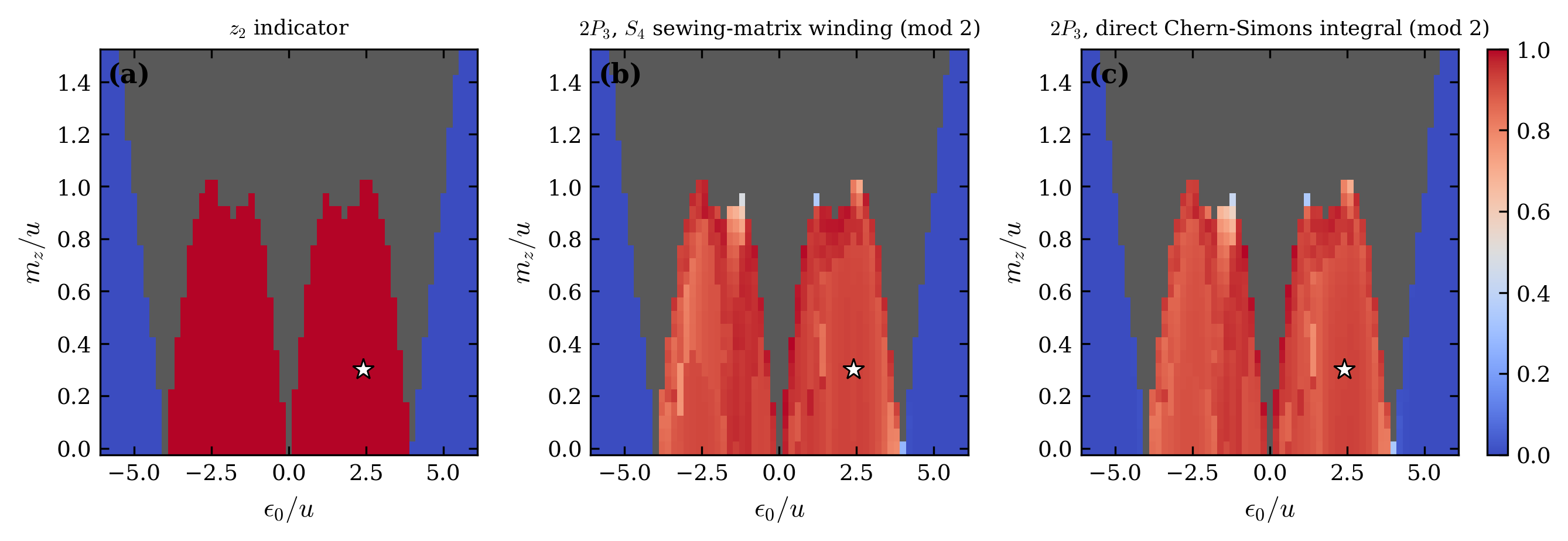}
    \caption{Comparison of the indicator and the axion invariant over the $(\epsilon_0,m_z)$ phase diagram of Fig.~\ref{fig:TB_phase}: (a) $z_2$ from the $S_4$ eigenvalues, (b) $2P_3$ from the mapping degree of the $S_4$ sewing matrix, Eq.~\eqref{eq:P3_with_B}, and (c) $2P_3$ from the direct Chern-Simons integral, Eq.~\eqref{P3_def}. In (b) and (c) the value is folded modulo 2 (distance from the nearest even integer), so $0$ is trivial and $1$ is the axion insulator; gapless points are grey.}
    \label{fig:TB_p3map}
\end{figure}

Figure~\ref{fig:TB_p3map} compares the $z_2$ indicator with both evaluations of $2P_3$: the mapping degree of $B_r(\bm k)$ in Eq.~\eqref{eq:P3_with_B} and the direct Chern-Simons integral in Eq.~\eqref{P3_def}. Each numerical value of $2P_3$ is plotted as its distance from the nearest even integer.
Among the 1009 insulating points, both evaluations agree with $z_2$ at every point where the numerical value is within $0.3$ of an integer. The only unresolved points (4 for the mapping degree and 7 for the direct integral) lie in cells adjacent to the phase boundary, where the gap is below $0.16u$ and the finite-difference integral has not converged even at $n_k=64$. At these points, the value drifts from $1$ towards $1/2$ but never to $0$.
The two evaluations of $2P_3$ also track each other cell by cell, confirming Eq.~\eqref{eq:P3_with_B} independently of the indicator.

\section{\label{sec:conclusion}Conclusion}
The equivalence between the Chern-Simons axion invariant $2P_3$ and the symmetry indicator $z_2$ has been established for $S_4$-symmetric axion insulators with vanishing three-dimensional Chern numbers.
The proof uses the $S_4$ sewing matrix, stable reblocking into $SU(2)$ two-band sectors, and the modulo-two degree of the resulting maps to $SU(2)$.
Although the modulo-two mapping-degree argument itself is not tied to four-fold symmetry, the explicit stable reduction and gauge fixing employed here rely on the even-order rotation subgroup $C_2=S_4^2$. Extending the construction to other rotoinversions such as $S_6$, whose rotation subgroup $C_3$ has odd order, would require a different treatment of the $U(2)$-to-$SU(2)$ reduction and is left for future work.
The methods developed here are also useful for connecting TBT and TFT for similar unitary symmetries.

A four-band tight-binding model whose magnetic point group is exactly $\bar{4}$ verifies the result by direct evaluation of the symmetry data, the sewing-matrix mapping degree, and the Chern-Simons integral, both at a single parameter point and at every insulating point of its exact phase diagram.
The model also illustrates how symmetry-enforced Weyl points connect the two occupied bands into the two-band blocks used in the proof.

The equivalence connects the response-theory and symmetry-indicator descriptions of $S_4$-protected axion phases.
In this sense, the symmetry indicator is given a direct field-theoretic meaning: it computes the quantized magnetoelectric response rather than only classifying high-symmetry band data.
The construction also shows how the response-indicator correspondence can survive when the quantizing symmetry is unitary and orientation reversing, broadening the link between topological field theory and topological band theory beyond the more familiar time-reversal or $C_nT$ cases.

\ack{We thank Zhi-Da Song for guidance and insightful discussions throughout this work.}

\clearpage
\appendix

\section{\label{app:sec1}$2P_3$ under $S_4$ symmetry}

The magnetoelectric polarization $P_3$ is formally given by the Chern-Simons integral
\begin{equation}
P_3 = \frac{1}{16\pi^2}\int d^3\bm{k} \epsilon^{ijk}
\text{Tr}\left[\left(\mathcal{F}^{ij}(\bm{k}) - \frac{2}{3}\text{i} \mathcal{A}^i(\bm{k}) \mathcal{A}^j(\bm{k})\right) \mathcal{A}^k(\bm{k})\right]
\end{equation}
where $\mathcal{A}^i_{mn}(\bm{k}) = -\text{i}\langle u_m(\bm{k})|\partial_{k_i}|u_n(\bm{k})\rangle$ is the non-Abelian Berry connection 
and $\mathcal{F}^{ij}(\bm{k}) = \partial_{k_i}\mathcal{A}^j(\bm{k}) - \partial_{k_j}\mathcal{A}^i(\bm{k}) + \text{i}[\mathcal{A}^i(\bm{k}),\mathcal{A}^j(\bm{k})]$ is the non-Abelian Berry curvature. The indices $m,n$ run over all occupied bands.
As discussed in Ref.~\cite{qi2008} and in the Supplemental Material of Ref.~\cite{chen2025}, $P_3$ is gauge invariant modulo 1 under transformations of the form $|u_n(\bm{k})\rangle\to|u_l(\bm{k})\rangle G_{ln}(\bm{k})$. 

The polarization $P_3$ can be expressed in terms of the sewing matrix $B(\bm{k})$. 
The strategy is analogous to that in the Supplemental Material of Ref.~\cite{chen2025}, but differs because $S_4$ is unitary.
First consider the transformation properties under $S_4\equiv C_4\cdot M_{xy}$, where $C_4$ is the four-fold rotation about the $z$ axis and $M_{xy}$ is a mirror reflection across the $xy$ plane.
Because the $S_4$ operator maps $|u_n(\bm{k})\rangle$ into a linear combination of occupied states at $S_4\bm{k}$, $B(\bm{k})$ is strictly unitary. 
The following relations hold:
\begin{equation}
    \begin{cases}
        S_4|u_n(\bm{k})\rangle = |u_m(S_4\bm{k})\rangle B_{mn}(\bm{k})\\
        |u_m(S_4\bm{k})\rangle = B^*_{mn}(\bm{k}) S_4|u_n(\bm{k})\rangle
    \end{cases}
\label{B_relation}
\end{equation}

Define the matrix $R$ representing the $S_4$ point-group operation in momentum space, $S_4\bm{k}=\bm{k'}$, such that $k_i'=R_{ij}k_j$. This transformation belongs to $O(3)$ and satisfies $RR^T=I$ and $\det(R)=-1$. 

Consequently, the Berry connection transforms as
\begin{equation}
\tilde{\mathcal{A}}^i(\bm{k'}) = - \text{i} R_{ij} B(\bm{k})\partial_{k_j}B^\dagger(\bm{k}) + R_{ij} B(\bm{k})\mathcal{A}^j(\bm{k}) B^\dagger(\bm{k})
\label{A_transform}
\end{equation}
and similarly for the Berry curvature (abbreviating $\partial_{k_i}$ as $\partial_i$ for conciseness):
\begin{equation}
\tilde{\mathcal{F}}^{ij}(\bm{k'}) = R_{i\alpha}R_{j\beta} B(\bm{k}) \mathcal{F}^{\alpha\beta}(\bm{k}) B^\dagger(\bm{k})
\end{equation}
where the identity $B\partial_\alpha B^\dagger B = B\partial_\alpha(B^\dagger B) - B B^\dagger \partial_\alpha B = -\partial_\alpha B$ has been used.

$P_3$ is invariant under the $S_4$ transformation, so

\begin{widetext}
\begin{equation}
\begin{aligned}
P_3 &= \frac{1}{16\pi^2}\int d^3\bm{k} \epsilon^{ijk} \text{Tr}\left\{\left(\mathcal{F}^{ij}(\bm{k}) - \frac{2}{3}\text{i} \mathcal{A}^i(\bm{k}) \mathcal{A}^j(\bm{k})\right) \mathcal{A}^k(\bm{k})\right\}\\
&= \frac{1}{16\pi^2}\int d^3\bm{k'} \epsilon^{ijk} \text{Tr}\left\{\left(\tilde{\mathcal{F}}^{ij}(\bm{k'}) - \frac{2}{3}\text{i} \tilde{\mathcal{A}}^i(\bm{k'}) \tilde{\mathcal{A}}^j(\bm{k'})\right) \tilde{\mathcal{A}}^k(\bm{k'})\right\}\\
&= \frac{1}{16\pi^2}\int d^3\bm{k} \epsilon^{ijk} R_{i\alpha}R_{j\beta}R_{k\gamma} \text{Tr}\{\left[B\mathcal{F}^{\alpha\beta}B^\dagger-\frac23\text{i}\left(B\mathcal{A}^\alpha B^\dagger-\text{i}B\partial_\alpha B^\dagger\right)\left( B\mathcal{A}^\beta B^\dagger-\text{i}B\partial_\beta B^\dagger\right)\right]\\
&\quad \left(B\mathcal{A}^\gamma B^\dagger-\text{i}B\partial_\gamma B^\dagger\right)\}
\end{aligned}
\end{equation}
\end{widetext}

Noting that $\epsilon^{ijk}R_{i\alpha}R_{j\beta}R_{k\gamma} = \det(R)\epsilon^{\alpha\beta\gamma} = -\epsilon^{\alpha\beta\gamma}$, we relabel the dummy indices $\alpha,\beta,\gamma$ as $i,j,k$. Using the cyclic property of the trace, $\text{Tr}[ABC]=\text{Tr}[BCA]=\text{Tr}[CAB]$, and $\text{Tr}[ABA^\dagger]=\text{Tr}[B]$ for unitary $A$, we obtain

\begin{widetext}
\begin{equation}
    \begin{aligned}
        P_3 &= -P_3 - \frac{1}{16\pi^2}\int d^3\bm{k} \epsilon^{ijk} \text{Tr}\{\text{i}\mathcal{F}^{ij}B^\dagger\partial_kB+2\mathcal{A}^i\mathcal{A}^jB^\dagger\partial_kB + 2\text{i}\mathcal{A}^i(B^\dagger\partial_jB)(B^\dagger\partial_kB) \\
        &\quad + \frac23(B^\dagger\partial_iB)(B^\dagger\partial_jB)(B^\dagger\partial_kB)\}
    \end{aligned}
\end{equation}
\end{widetext}

Expanding $\mathcal{F}^{ij}$ and using the total antisymmetry of $\epsilon^{ijk}$ gives

\begin{widetext}
\begin{equation}
\begin{aligned}
2P_3 &= - \frac{1}{16\pi^2}\int d^3\bm{k} \epsilon^{ijk} \text{Tr}\{(\mathcal{A}^i\mathcal{A}^j+\mathcal{A}^j\mathcal{A}^i)B^\dagger\partial_kB-2\text{i}(\partial_j\mathcal{A}^i)B^\dagger\partial_kB-2\text{i}\mathcal{A}^i\partial_jB^\dagger\partial_kB\\
&\quad + \frac23(B^\dagger\partial_iB)(B^\dagger\partial_jB)(B^\dagger\partial_kB)\}\\
&= -\frac{1}{16\pi^2}\int d^3\bm{k} \epsilon^{ijk} \text{Tr}\left[-2\text{i}(\partial_j\mathcal{A}^i)B^\dagger\partial_kB-2\text{i}\mathcal{A}^i\partial_jB^\dagger\partial_kB+\frac23(B^\dagger\partial_iB)(B^\dagger\partial_jB)(B^\dagger\partial_kB)\right]\\
&= -\frac{1}{16\pi^2}\int d^3\bm{k} \epsilon^{ijk} \text{Tr}\left[-2\text{i}\partial_j(\mathcal{A}^iB^\dagger\partial_kB)+\frac23(B^\dagger\partial_iB)(B^\dagger\partial_jB)(B^\dagger\partial_kB)\right]\\
&= -\frac{1}{24\pi^2}\int d^3\bm{k} \epsilon^{ijk} \text{Tr}\left[(B^\dagger\partial_iB)(B^\dagger\partial_jB)(B^\dagger\partial_kB)\right]
\end{aligned}
\label{eq:P3_B}
\end{equation}
\end{widetext}

This expression holds modulo 2, irrespective of whether the operator convention gives $(S_4)^4=-1$ or $(S_4)^4=1$.
A smooth $B(\bm{k})$ is required to define the differentials in Eq.~\eqref{eq:P3_B}. 
As explained in the Supplemental Material of Ref.~\cite{chen2025}, this condition can be satisfied when the system has vanishing three-dimensional Chern numbers, as assumed here. 
For topological states protected by symmetries other than translation, a symmetry-breaking Wannier gauge can be chosen~\cite{soluyanov2011} in which the states are smooth over the Brillouin zone.

\section{\label{app:SU2_example}An explicit example of reduction from $U(2)$ to $SU(2)$ by adding atomic bands}

As an explicit example of the stable equivalence, consider the two-band system with $S_4$ eigenvalues:
\begin{equation}
\begin{aligned}
&\Gamma: e^{i\frac{\pi}{4}},e^{i\frac{\pi}{4}} &&A: e^{-i\frac{\pi}{4}},e^{-i\frac{\pi}{4}}\\
&M: e^{i\frac34\pi},e^{-i\frac{\pi}{4}} &&Z: e^{-i\frac34\pi},e^{i\frac{\pi}{4}}
\end{aligned}
\end{equation}
and $C_2$ eigenvalues:
\begin{equation}
\begin{aligned}
&R: e^{i\frac{\pi}{2}},e^{i\frac{\pi}{2}} &&X: e^{i\frac{\pi}{2}},e^{i\frac{\pi}{2}}
\end{aligned}
\end{equation}
The corresponding symmetry-data vector is
\begin{equation}
\mathbf b=\,2\bar\Gamma_8+2\bar A_6+\bar M_5+\bar M_6+\bar Z_7+\bar Z_8+2\bar R_4+2\bar X_4 ,
\end{equation}
which has $(z_{4S},\delta_{2S},z_2)=(0,0,1)$ but is not directly $SU(2)$-reducible. 
We add the following three EBRs:
\begin{equation}
    \text{EBR 1:}
    \begin{cases}
    \begin{aligned}
        &\Gamma: e^{-i\frac{\pi}{4}} &&A: e^{i\frac34\pi} &&M: e^{-i\frac{\pi}{4}}\\
        &Z: e^{i\frac34\pi} &&R: e^{-i\frac{\pi}{2}} &&X: e^{-i\frac{\pi}{2}}
    \end{aligned}
    \end{cases}
\end{equation}
\begin{equation}
    \text{EBR 2:}
    \begin{cases}
    \begin{aligned}
        &\Gamma: e^{-i\frac34\pi} &&A: e^{i\frac{\pi}{4}} &&M: e^{i\frac{\pi}{4}}\\
        &Z: e^{-i\frac34\pi} &&R: e^{-i\frac{\pi}{2}} &&X: e^{-i\frac{\pi}{2}}
    \end{aligned}
    \end{cases}
\end{equation}
\begin{equation}
    \text{EBR 3:}
    \begin{cases}
    \begin{aligned}
        &\Gamma: e^{i\frac34\pi},e^{-i\frac{\pi}{4}} &&A: e^{-i\frac34\pi},e^{i\frac{\pi}{4}} &&M: e^{-i\frac34\pi},e^{i\frac{\pi}{4}}\\
        &Z: e^{i\frac34\pi},e^{-i\frac{\pi}{4}} &&R: e^{-i\frac{\pi}{2}},e^{i\frac{\pi}{2}} &&X: e^{-i\frac{\pi}{2}},e^{i\frac{\pi}{2}}
    \end{aligned}
    \end{cases}
\end{equation}
The resulting symmetry-data vector can be decomposed into three determinant-one two-band generators of the form defined in Eq.~\eqref{eq:det1-two-band-vector}, as shown in Table~\ref{tab:det1-decomposition}.
\begin{table}[htbp]
    \centering
    \begin{tabular}{|c|c|c|c|}
        \hline
        & Bands 1\&2 & Bands 3\&4 & Bands 5\&6\\\hline
        $\Gamma$ & $e^{\pm i\frac{\pi}{4}}$ & $e^{\pm i\frac{\pi}{4}}$ & $e^{\pm i\frac34\pi}$\\\hline
        $A$ & $e^{\pm i\frac{\pi}{4}}$ & $e^{\pm i\frac34\pi}$ & $e^{\pm i\frac{\pi}{4}}$\\\hline
        $M$ & $\quad e^{\pm i\frac34\pi}$ & $e^{\pm i\frac{\pi}{4}}$ & $e^{\pm i\frac{\pi}{4}}$\\\hline
        $Z$ & $e^{\pm i\frac{\pi}{4}}$ & $e^{\pm i\frac34\pi}$ & $e^{\pm i\frac34\pi}$\\\hline
        $R$ & $e^{\pm i\frac{\pi}{2}}$ & $e^{\pm i\frac{\pi}{2}}$ & $e^{\pm i\frac{\pi}{2}}$\\\hline
        $X$ & $e^{\pm i\frac{\pi}{2}}$ & $e^{\pm i\frac{\pi}{2}}$ & $e^{\pm i\frac{\pi}{2}}$\\\hline
    \end{tabular}
    \caption{Symmetry data of three determinant-one two-band generators.}
    \label{tab:det1-decomposition}
\end{table}
This illustrates the stable relation
\begin{equation}
\mathbf b+\mathcal E_{i_1}+\mathcal E_{i_2}+\mathcal E_{i_3}
=S_2^{(j_1)}+S_2^{(j_2)}+S_2^{(j_3)} ,
\end{equation}
Thus, after adding three EBRs, the six bands can be grouped into three two-band systems, each of which admits direct reduction to $SU(2)$.

\section{\label{app:sec4}Transforming $D_r(\bm{k})$ into constant matrices}
Following the construction in the Supplemental Material of Ref.~\cite{li2020}, we construct an explicit transformation that converts $D_r(\bm{k})$ into $\text{i}\sigma_z$.

Since $S_4^2=C_2$ and $(S_4)^4=-1$, the definition of $D(\bm{k})$ in Eq.~\eqref{D_def} gives
\begin{equation}
    D_{mn}(C_2\mathbf{k}) = \langle u_m(\mathbf{k})|C_2u_n(C_2\mathbf{k})\rangle = -\langle C_2u_m(\mathbf{k})|u_n(C_2\mathbf{k})\rangle = -D_{nm}^*(\mathbf{k})
\end{equation}
In other words:
\begin{equation}
    D(C_2\bm{k})=-D^\dagger(\bm{k})
    \label{eq:D_symmetry}
\end{equation}

Since $B(\bm{k})$ has been reduced to $SU(2)$, the corresponding $D(\bm{k})$ is also in $SU(2)$ and can be parameterized as
\begin{equation}
    D(\bm{k})=\exp[\text{i}(\pi/2+\delta(\bm{k}))\hat{n}(\bm{k})\cdot\vec{\sigma}]
\end{equation}
where $\delta(\bm{k})\in[-\frac{\pi}{2},\frac{\pi}{2}]$ and $\hat{n}(\bm{k})$ is a unit vector.
The symmetry requirement in Eq.~\eqref{eq:D_symmetry} implies
\begin{equation}
\delta(\bm{k})=-\delta(C_2\bm{k}),\quad \hat{n}(\bm{k})=\hat{n}(C_2\bm{k})
\label{eq:delta_n_symmetry}
\end{equation}

Under an $SU(2)$ gauge transformation $G(\bm{k})$, $D(\bm{k})\to G^\dagger(C_2\bm{k})D(\bm{k})G(\bm{k})$.
With the transformation $R(\bm{k})=\exp[-\text{i}\frac12\delta(\bm{k})\hat{n}(\bm{k})\cdot\vec{\sigma}]$, we have
\begin{equation}
    \begin{aligned}
        &R^\dagger(C_2\bm{k})D(\bm{k})R(\bm{k})\\
        =&\exp[-\text{i}\frac12\delta(\bm{k})\hat{n}(\bm{k})\cdot\vec{\sigma}]\exp[\text{i}(\pi/2+\delta(\bm{k}))\hat{n}(\bm{k})\cdot\vec{\sigma}]\\
        &\exp[-\text{i}\frac12\delta(\bm{k})\hat{n}(\bm{k})\cdot\vec{\sigma}]\\
        =&\exp[\text{i}\frac{\pi}{2}\hat{n}(\bm{k})\cdot\vec{\sigma}]=\text{i}\hat{n}(\bm{k})\cdot\vec{\sigma}
    \end{aligned}
\end{equation}
thereby removing the $\delta(\bm{k})$ dependence.

From the Supplemental Material of Ref.~\cite{li2020}, the degree of the map from the Brillouin zone to $D(\bm{k})$ satisfies
\begin{equation}
    \begin{cases}
        \deg[D(C_2\bm{k})]=\deg[D(\bm{k})]\\
        \deg[D^\dagger(\bm{k})]=-\deg[D(\bm{k})]
    \end{cases}
    \label{eq:degree_symmetry}
\end{equation}
Equations~\eqref{eq:D_symmetry} and \eqref{eq:degree_symmetry} imply that $\deg[D(\bm{k})]=0$.
Therefore there exists a similarity transformation $w(\bm{k})$ to bring each $\hat{n}(\bm{k})$ to $\hat{z}$, namely
\begin{equation}
w^\dagger(\bm{k})(\text{i}\hat{n}(\bm{k})\cdot\vec{\sigma})w(\bm{k})=\text{i}\sigma_z
\end{equation}
As a result, we also have:
\begin{equation}
w^\dagger(C_2\bm{k})(\text{i}\hat{n}(C_2\bm{k})\cdot\vec{\sigma})w(C_2\bm{k})=\text{i}\sigma_z
\end{equation}
With the residual gauge chosen to respect Eq.~\eqref{eq:delta_n_symmetry}, we take $w(C_2\bm{k})=w(\bm{k})$.

Then, the overall transformation $G(\bm{k})=R(\bm{k})w(\bm{k})$ brings $D(\bm{k})$ to $\text{i}\sigma_z$:
\begin{equation}
    \begin{aligned}
        &G^\dagger(C_2\bm{k})D(\bm{k})G(\bm{k})\\
        =&w^\dagger(\bm{k})R^\dagger(C_2\bm{k})D(\bm{k})R(\bm{k})w(\bm{k})\\
        =&w^\dagger(\bm{k})(\text{i}\hat{n}(\bm{k})\cdot\vec{\sigma})w(\bm{k})=\text{i}\sigma_z
    \end{aligned}
\end{equation}

Thus the degree of the map from the Brillouin zone to $D(\bm{k})$ vanishes, allowing a continuous deformation of $D_r(\bm{k})$ into a constant matrix without violating the symmetry requirement in Eq.~\eqref{eq:D_symmetry}.
This constant matrix is chosen to be $\text{i}\sigma_z$ for convenience.

\section{\label{app:sec5}Regularity of $A_\pm$}

We show that $A_\pm$ are not forced by symmetry to be singular values by analyzing the local expansion of $B_r(\bm{k})$ near the $S_4$-invariant momenta $K$. 
Here a target value is regular if, at every preimage $\bm{k}$ with $B_r(\bm{k})=A_\pm$, the Jacobian of the map from local coordinates of the Brillouin zone to local coordinates of $SU(2)$ has full rank.
These fixed momenta require special attention because they have the largest little group.
At a generic momentum, the four points in its $S_4$ orbit are distinct; a nonsingular local Jacobian at one point can be extended to the other three points by symmetry, so the symmetry imposes no local rank obstruction there.
At a momentum fixed only by $C_2=S_4^2$, the constraint is $B_r(C_2\bm k)=\sigma_z B_r(\bm k)\sigma_z$.
Near $A_+$, the domain tangent space splits as two $C_2$-odd directions $(p_x,p_y)$ and one $C_2$-even direction $p_z$, while the target coordinates $(d_x,d_y)$ are odd and $d_0$ is even.
Thus a full-rank equivariant differential is also allowed on the $C_2$-invariant lines.

Consider first $A_+=\frac{1}{\sqrt2}(\sigma_0+\text{i}\sigma_z)$.
An explicit local model satisfying the $SU(2)$ and $S_4$ constraints is
\begin{equation}
    \begin{aligned}
        B_+(\bm{p})&=\frac{\widetilde B_+(\bm{p})}{\sqrt{\det\widetilde B_+(\bm{p})}},\\
        \widetilde B_+(\bm{p})&=(\frac{1}{\sqrt{2}}+p_z)\sigma_0+\text{i}p_x\sigma_x-\text{i}p_y\sigma_y+\text{i}(\frac{1}{\sqrt2}-p_z)\sigma_z .
    \end{aligned}
\end{equation}
Here
\begin{equation}
\det\widetilde B_+(\bm{p})=1+p_x^2+p_y^2+2p_z^2,
\end{equation}
so $B_+(\bm p)\in SU(2)$ and the normalization does not change the linear term.
This local model satisfies the following requirements:
\begin{itemize}
    \item Fixed-point value: $B_+(0)=A_+$.
    \item Symmetry conjugation constraint:
    \begin{equation}
    B_+(p_y,-p_x,-p_z)=\text{i}\sigma_z B_+^\dagger(p_x,p_y,p_z).
    \end{equation}
\end{itemize}
Writing $B_+=d_0\sigma_0+\text{i}d_x\sigma_x+\text{i}d_y\sigma_y+\text{i}d_z\sigma_z$, the $SU(2)$ constraint fixes $d_z$ locally in terms of $d_0$, $d_x$, and $d_y$ because $d_z(A_+)=1/\sqrt2\neq0$.
Thus $(d_0,d_x,d_y)$ are valid local coordinates on the target $S^3$ near $A_+$.
The Jacobian at $\bm{p}=0$ is
\begin{equation}
    J=\frac{\partial(d_0,d_x,d_y)}{\partial(p_x,p_y,p_z)}
    =
    \begin{pmatrix}
    0&0&1\\
    1&0&0\\
    0&-1&0
    \end{pmatrix},
    \qquad
    \det J=-1.
\end{equation}
Equivalently, to first order,
\begin{equation}
    d_0=\frac{1}{\sqrt{2}}+p_z,\quad
    d_x=p_x,\quad
    d_y=-p_y,\quad
    d_z=\frac{1}{\sqrt2}-p_z .
\end{equation}
The determinant is nonzero, so $A_+$ is a regular value for this symmetry-allowed local model at the fixed momentum.
Therefore $A_+$ is not symmetry-forced to be a singular value.
If a particular $B_r$ has a singular differential at such a preimage, that singularity is accidental and can be removed by an arbitrarily small symmetry-preserving perturbation of the local map while keeping the fixed-point value $A_+$ unchanged.
The case $A_-$ follows by taking the local model $B_-(\bm p)=-B_+(\bm p)$.

\section{\label{app:TB_details}Details of the tight-binding model}

\subsection{\label{subsec:four_band_model}\label{app:TB_symmetry_audit}Spectrum and symmetry analysis}

\begin{figure}[htbp]
    \centering
    \includegraphics[width=0.75\textwidth]{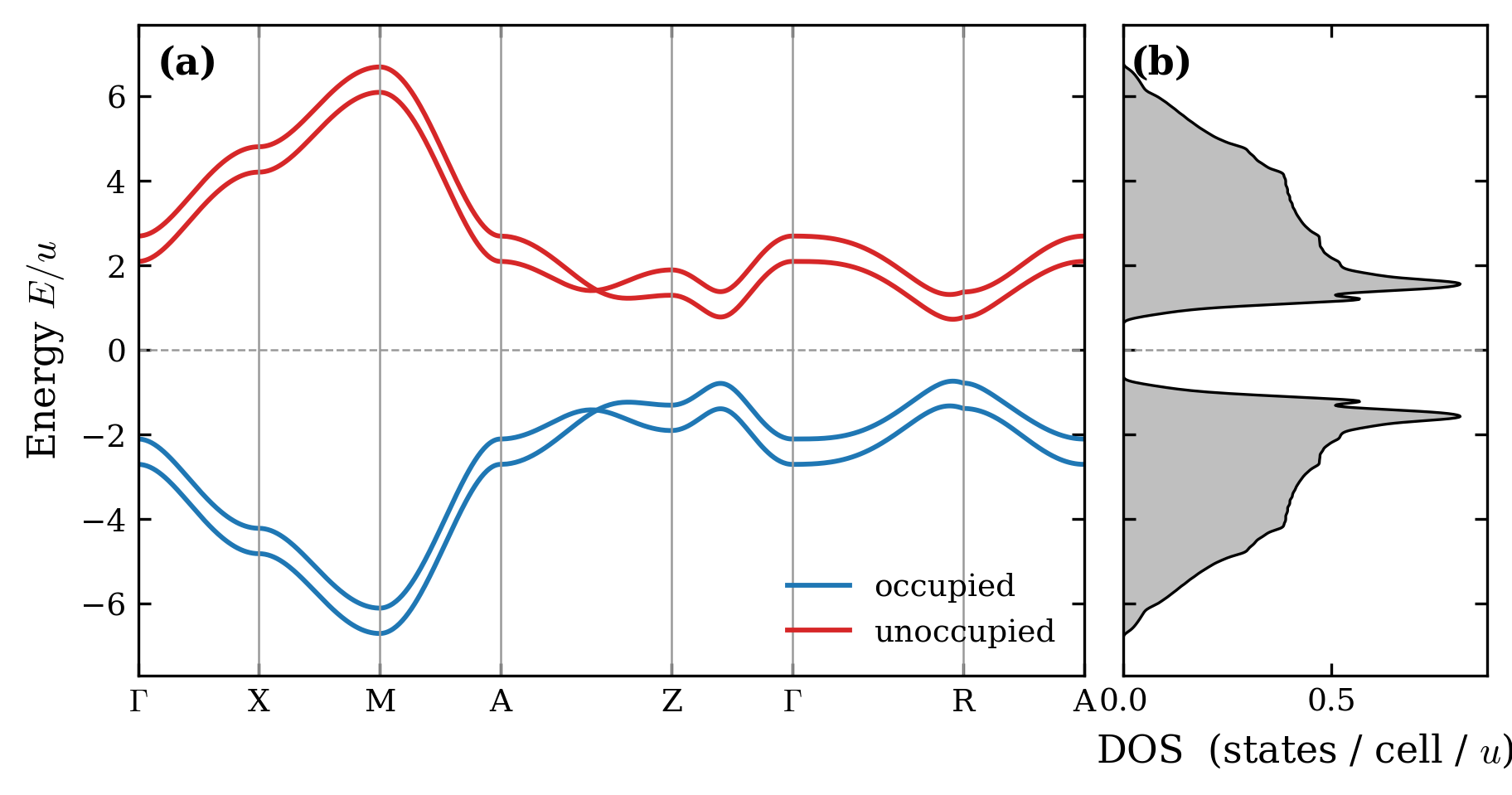}
    \caption{(a) Band structure of the model of Eq.~\eqref{eq:TB_model} along $\Gamma$--X--M--A--Z--$\Gamma$--R--A and (b) density of states, for $u=v=1$, $\epsilon_0=2.4$, $m_z=0.3$, $\eta=\kappa=0.5$. The two occupied bands are drawn in blue; they touch at four Weyl points on the plane $k_z=\pi$ (not on the path shown) but are separated from the unoccupied bands by a gap of $1.44u$.}
    \label{fig:TB_band}
\end{figure}

The spectrum is available in closed form:
\begin{eqnarray}
E(\bm{k})&&=\pm\sqrt{\big(\rho(\bm{k})\pm m_z\big)^2+\sin^2k_x+\sin^2k_y},\nonumber\\
\rho(\bm{k})&&\equiv\sqrt{\epsilon^2+\sin^2k_z+g^2}.
\label{eq:TB_spectrum}
\end{eqnarray}
The band structure and density of states are shown in Fig.~\ref{fig:TB_band}; the minimum bulk gap is $1.44u$.

In this subsection, we verify the magnetic point group of the tight-binding model in Eq.~\eqref{eq:TB_model} by direct computation.
From the full cubic holohedry $\mathcal O_h$, we find that the only symmetries of the model are $\{E, C_{2z}, S_{4z}, S_{4z}^{-1}\}$, so the results of the main text derive from the $S_4$ symmetry alone. 
Although the model is defined on a tetragonal lattice, we test symmetries beyond the tetragonal holohedry $\mathcal D_{4h}$ to ensure that no accidental symmetries are present.
We use the same model parameters as in Fig.~\ref{fig:TB_band}.

Every symmorphic point operation is represented in momentum space by one of the 48 signed permutation matrices $R\in \mathcal O_h$.
We test each of these operations on the model, both without and with time reversal, giving the 96 candidates in $\mathcal O_h\times\{1,T\}$.
For a candidate to be admissible, a constant $4\times4$ matrix $U$ must satisfy, respectively,
\begin{equation}
\begin{aligned}
H(R\bm k)U&=UH(\bm k) &&\text{(unitary)},\\
H(R\bm k)U&=UH^*(-\bm k) &&\text{(antiunitary)}.
\end{aligned}
\label{eq:TB_symmetry_test}
\end{equation}
Here $R\in\mathcal{O}_h$ denotes the action of the candidate operation on momentum in either line.

Equation~\eqref{eq:TB_symmetry_test} is a linear system in the 16 complex entries of $U$.
We stack the equations evaluated at 40 generic random momenta and compute the common null space using singular value decomposition (SVD).
A candidate is accepted only when this null space is nonzero.
Only four candidates have nonzero null spaces; each null space is one dimensional, and its normalized generator is unitary with a maximum error below $2\times10^{-15}$:
\begin{equation}
\begin{array}{c|c}
\text{spatial operation} & \text{internal representation}\\ \hline
E & \tau_0\otimes\sigma_0\\
C_{2z} & U_{S_4}^2=-\text{i}\,\tau_0\otimes\sigma_z\\
S_4 & U_{S_4}\\
S_4^{-1} & U_{S_4}^{-1}.
\end{array}
\end{equation}
No antiunitary candidate has a nonzero null space.
The magnetic point group of the model is therefore exactly
\begin{equation}
\{E,C_{2z},S_4,S_4^{-1}\}=\bar 4.
\end{equation}

\subsection{\label{subsec:symmetry_data}Symmetry indicators and $2P_3$ from preimage counting}
At an $S_4$-invariant momentum $K$, all sine terms and $g$ vanish, so $H(K)=\epsilon(K)\tau_x+m_z\sigma_z$.
For $|\epsilon(K)|>m_z$ the two occupied states have $\tau_x=-\text{sgn}\,\epsilon(K),\,\sigma_z=\pm1$, and are split by $2m_z$, carrying the $S_4$ eigenvalues $-\text{sgn}\,\epsilon(K)\,e^{\mp\text{i}\pi/4}$.
With $\epsilon(\Gamma)=\epsilon(A)=\epsilon_0$, $\epsilon(Z)=\epsilon_0-4$, and $\epsilon(M)=\epsilon_0+4$, the occupied pair has eigenvalues $e^{\pm\text{i}\pi/4}$ at $Z$ and $e^{\pm\text{i}3\pi/4}$ at $\Gamma$, $M$, and $A$, while the $C_2$ eigenvalues at $R$ (and $X$) are $\pm\text{i}$.
Hence $\{z_{4S},\delta_{2S},z_2\}=\{0,0,1\}$.
For the two-band occupied subspace, all six Chern numbers on the planes $k_i=0,\pi$ vanish, as computed on a $40\times40$ momentum mesh using the lattice Berry-curvature method of Ref.~\cite{fukui2005}.
In the language of Sec.~\ref{sec:winding_indicator}, the occupied $2\times2$ sewing block at the $S_4$-invariant momenta is
\begin{eqnarray}
D(K)&&=\text{i}\sigma_z,\qquad B(Z)=\begin{bmatrix}e^{\text{i}\frac{\pi}{4}}&\\&e^{-\text{i}\frac{\pi}{4}}\end{bmatrix}=A_+,\nonumber\\
B(\Gamma,M,A)&&=\begin{bmatrix}-e^{\text{i}\frac{\pi}{4}}&\\&-e^{-\text{i}\frac{\pi}{4}}\end{bmatrix}=A_-,
\end{eqnarray}
so the preimage count of $A_+$ over $K^4$ is $n_r=1$ and Sec.~\ref{sec:winding_indicator} predicts $2P_3=1$.
More generally, the same argument shows that in this model $z_2$ equals the number of $S_4$-invariant momenta with $\epsilon(K)>0$ modulo 2, which is $1$ for $0<|\epsilon_0|<4$ and $0$ for $|\epsilon_0|>4$.
In other words, a topological axion-insulator phase can occur only for $0<|\epsilon_0|<4$, whereas the phase is trivial for $|\epsilon_0|>4$.

\subsection{\label{subsec:weyl_points}Weyl points inside the occupied space}

In this subsection we show that the two occupied bands are connected by Weyl points.
The presence of these Weyl points makes the $2\times2$ block irreducible, consistent with the general argument in Sec.~\ref{sec:block_decomposition} that sewing-matrix blocks of at least two bands are needed to realize the $S_4$ axion phase.

The two occupied bands are not isolated from each other.
By Eq.~\eqref{eq:TB_spectrum}, $E_1=E_2$ requires $\rho=0$, i.e., $\sin k_z=0$, $\cos k_x=\cos k_y$, and $\epsilon=0$, which for $0<\epsilon_0<4$ has exactly four solutions
\begin{equation}
    \bm{k}^{W}=(\pm k^*,\pm k^*,\pi),\qquad \cos k^*=\frac{\epsilon_0-2}{2},
\end{equation}
with both signs independent ($k^*=0.436\pi$ for $\epsilon_0=2.4$; for $-4<\epsilon_0<0$ the four nodes lie on $k_z=0$ instead, and for $|\epsilon_0|>4$ there are none).
The four nodes form a single $S_4$ orbit and are Weyl points of band~1 with alternating chirality $+1,-1,+1,-1$, obtained as the Berry flux of the lowest band through a small closed cube around each node.
These crossings obstruct globally smooth eigenstates for the individual occupied bands, i.e., the sewing-matrix blocks cannot be diagonalized globally.
The irreducibility of the occupied $2\times2$ sewing matrix block follows directly from $2P_3=1$: a decomposition into two $1\times1$ blocks in a globally smooth, periodic gauge would give zero in Eq.~\eqref{eq:P3_with_B}, as discussed in Sec.~\ref{sec:block_decomposition}.
The nodes do not obstruct the axion invariant, since the occupied projector $P_{\rm occ}(\bm{k})=\sum_{n\in\rm occ}|u_n(\bm{k})\rangle\langle u_n(\bm{k})|$ remains smooth as long as the gap to band~3 stays open.

\subsection{\label{subsec:2P_3_integral}$2P_3$ from direct integration}

$2P_3$ was evaluated in Appendix~\ref{subsec:symmetry_data} by counting the preimages $n_r$.
We now evaluate $2P_3$ directly using Eq.~\eqref{eq:P3_with_B} and obtain the same result.

At each point of the momentum grid, we diagonalize the $4\times4$ Hamiltonian and collect the two lowest-energy eigenvectors in $U_{\rm occ}(\bm{k})=[|u_1(\bm{k})\rangle,|u_2(\bm{k})\rangle]$.
These eigenvectors carry an arbitrary $U(2)$ gauge at every $\bm{k}$, making the numerical evaluation of $B=U_{\rm occ}^\dagger(S_4\bm k)U_{S_4}U_{\rm occ}(\bm k)$ and its derivatives in Eq.~\eqref{eq:P3_with_B} unstable.
Numerically evaluating Eq.~\eqref{eq:P3_with_B} requires a smooth, periodic, orthonormal frame $V_{\rm occ}(\bm{k})=[|v_1(\bm{k})\rangle,|v_2(\bm{k})\rangle]$, where $|v_{1,2}(\bm k)\rangle$ are obtained from $|u_{1,2}(\bm k)\rangle$ by a unitary transformation within the occupied subspace.
We can then rewrite $B(\bm k)$ as
\begin{equation}
B(\bm{k})=V_{\rm occ}^\dagger(S_4\bm{k})U_{S_4}V_{\rm occ}(\bm{k})
\end{equation}
and use it to evaluate Eq.~\eqref{eq:P3_with_B} stably. Below, we describe how to construct such a smooth frame $V_{\rm occ}(\bm{k})$.

Although the eigenvectors $|u_{1,2}(\bm k)\rangle$ have an arbitrary gauge, their combined projector
\begin{equation}
P_{\rm occ}(\bm{k})=U_{\rm occ}(\bm{k})U_{\rm occ}^\dagger(\bm{k})
=\sum_{n=1}^{2}|u_n(\bm{k})\rangle\langle u_n(\bm{k})|
\end{equation}
is gauge invariant and remains smooth as long as the gap between the occupied and unoccupied bands remains open, even when the two occupied bands cross.
Therefore, we can construct a smooth frame $V_{\rm occ}(\bm{k})$ from $P_{\rm occ}(\bm{k})$.
For instance, we can choose two random vectors $R=[r_1,r_2]\in\mathbb{C}^{4\times2}$, project them onto the occupied subspace, and orthonormalize the result to obtain
\begin{equation}
V_{\rm occ}(\bm{k})=P_{\rm occ}(\bm{k})R(R^\dagger P_{\rm occ}(\bm{k})R)^{-1/2}.
\end{equation}
The inverse square-root factor ensures that $V_{\rm occ}^\dagger(\bm{k})V_{\rm occ}(\bm{k})=I_2$.
However, the matrix $R^\dagger P_{\rm occ}(\bm{k})R$ can become nearly singular along curves in the Brillouin zone, so we use the more robust construction of $V_{\rm occ}(\bm{k})$ described below.

The first vector can be obtained reliably as $|v_1(\bm{k})\rangle=P_{\rm occ}(\bm{k})r_1/\|P_{\rm occ}(\bm{k})r_1\|$, where $r_1\in\mathbb{C}^4$ is a fixed random vector.
In practice, several choices of $r_1$ are tested, and we select the one that maximizes $\min_{\bm{k}}\|P_{\rm occ}(\bm{k})r_1\|$.
The second vector spans the complementary line bundle $L=\mathrm{occ}\ominus v_1$, whose projector is $P_L=P_{\rm occ}-|v_1\rangle\langle v_1|$.
For a fixed random vector $r_2\in\mathbb{C}^4$, the projection $P_Lr_2$ may be small at some $\bm{k}$, leading to poor convergence of $2P_3$. 
Below, we describe a parallel-transport method for constructing $|v_2(\bm{k})\rangle$ as a function of momentum that avoids this problem.

At the grid point $(-\pi,-\pi,-\pi)$, we choose a random vector $r_2$ and initialize $|v_2\rangle=P_Lr_2/\|P_Lr_2\|$.
Along a grid direction, the section is parallel transported by successive projection and normalization,
\begin{equation}
|\widetilde v_{2,j}\rangle=\frac{P_L(\bm{k}_j)|\widetilde v_{2,j-1}\rangle}
{\|P_L(\bm{k}_j)|\widetilde v_{2,j-1}\rangle\|},\qquad j=1,\ldots,n_k-1.
\end{equation}
This operation is first performed along the $k_x$ line at $k_y=k_z=-\pi$, then along $k_y$ for every $k_x$ on the plane $k_z=-\pi$, and finally along $k_z$ for every $(k_x,k_y)$ in the bulk.
Constructing $|\widetilde v_{2,j}\rangle$ on the momentum grid in this way ensures that the norm of $P_L(\bm{k}_j)|\widetilde v_{2,j-1}\rangle$ remains sufficiently large throughout the Brillouin zone.
However, there is a phase mismatch between the first and last points along each transport line. 
These two points are neighbors in the periodic Brillouin zone, but $|\widetilde v_{2,0}\rangle$ and $|\widetilde v_{2,n_k-1}\rangle$ may differ by a phase factor due to the parallel transport:
\begin{equation}
\phi(\bm{k}_\perp)=\arg\!\left[\langle\widetilde v_{2,0}|
\frac{P_L(\bm{k}_0)|\widetilde v_{2,n_k-1}\rangle}
{\|P_L(\bm{k}_0)|\widetilde v_{2,n_k-1}\rangle\|}\right],
\end{equation}
where $\bm{k}_0$ and $\bm{k}_{n_k-1}$ are the first and last grid points along the transport direction, and $\bm{k}_\perp$ denotes the transverse momenta.
We first unwrap $\phi$ as a smooth function of $\bm{k}_\perp$. We then distribute the phase mismatch along the line by applying a phase factor at each grid point:
\begin{equation}
|v_{2,j}\rangle=\exp\!\left(-\mathrm{i}\phi(\bm{k}_\perp)\frac{j}{n_k}\right)|\widetilde v_{2,j}\rangle,
\end{equation}
where $k_j=-\pi+2\pi j/n_k$ denotes the coordinate along the transported direction.
This distributes the mismatch uniformly.
The vanishing Chern numbers allow the phase to be unwrapped without winding.

With $V_{\rm occ}(\bm{k})=[|v_1(\bm{k})\rangle,|v_2(\bm{k})\rangle]$ on an $n_k^3$ grid, the sewing matrix is $B(\bm{k})=V_{\rm occ}^\dagger(S_4\bm{k})U_{S_4}V_{\rm occ}(\bm{k})$ and Eq.~\eqref{eq:P3_with_B} is evaluated with periodic central differences.
For $n_k=24$, $32$, $48$ we obtain $2P_3=0.925$, $0.957$, $0.980$, converging to $1$ with the expected $n_k^{-2}$ discretization error.
As an independent check that does not use $S_4$, the Chern-Simons integral of Eq.~\eqref{P3_def} was evaluated directly with the same frame and gives identical results on each grid.
Finally, $2P_3=0$ is obtained in the same way for $\epsilon_0=\pm5$, where the indicator vanishes.

\subsection{\label{subsec:phase_diagram}Phase diagram and comparison with $z_2$}
The model in Eq.~\eqref{eq:TB_model} can be tuned to realize different phases beyond the parameter setting of Fig.~\ref{fig:TB_band}, for which the equivalence was tested in Appendices~\ref{subsec:symmetry_data} and \ref{subsec:2P_3_integral}.
In this subsection, we investigate the band gaps, the $z_2$ indicator obtained from $S_4$ eigenvalues, and $2P_3$ evaluated using both the integral expression for the mapping degree in Eq.~\eqref{eq:P3_with_B} and the direct Chern-Simons integral in Eq.~\eqref{P3_def} over a wide range of parameters.
The results are consistent, with $2P_3$ agreeing with $z_2$ at every insulating point, and the phase boundaries are determined by gap closings.

\begin{figure}[htbp]
    \centering
    \includegraphics[width=0.8\textwidth]{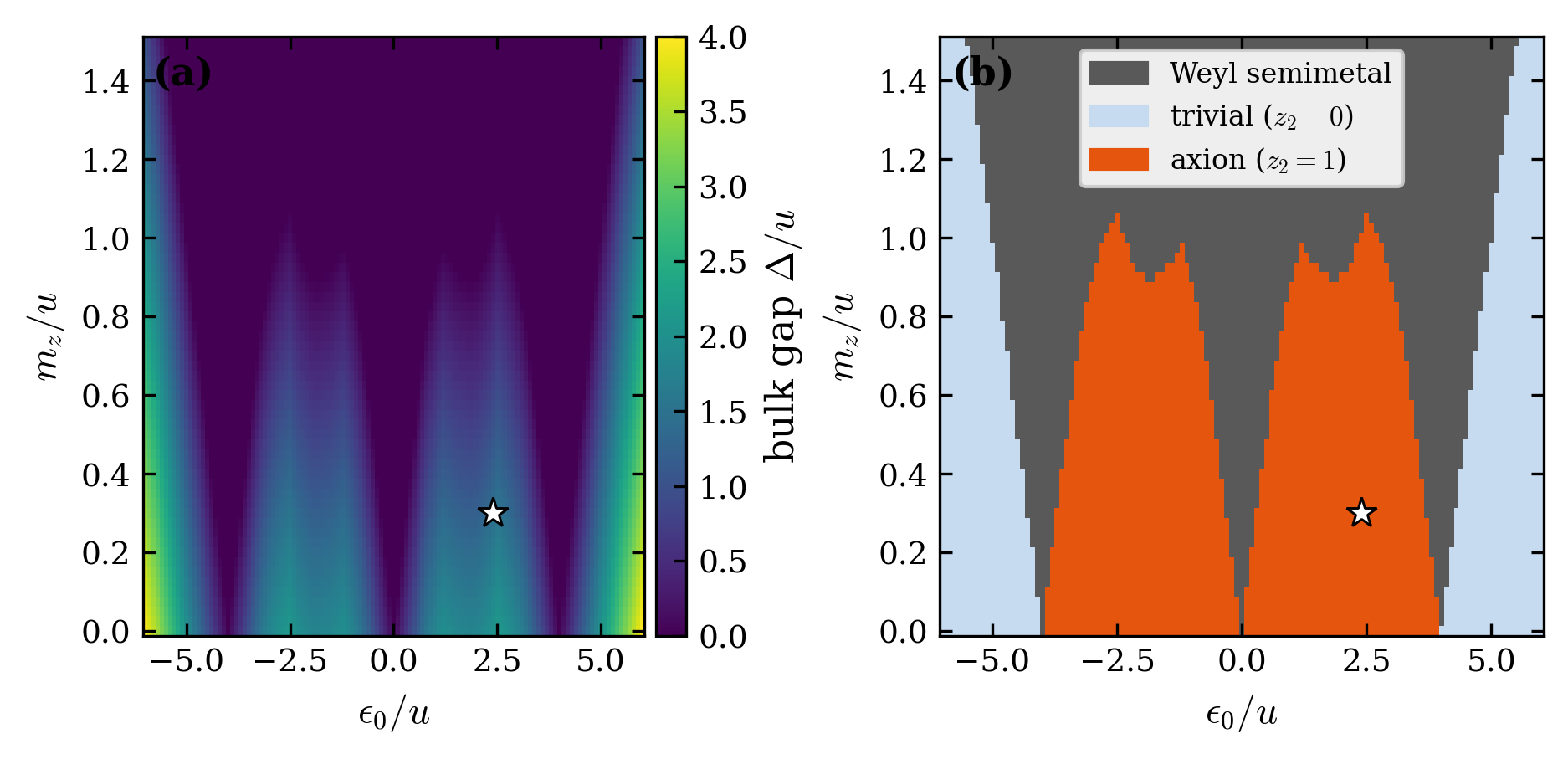}
    \caption{Phase diagram of the model in the $(\epsilon_0,m_z)$ plane for $\eta=\kappa=0.5$: (a) bulk gap $E_3-E_2$ and (b) phases determined from the symmetry indicators. The star marks the parameters of Fig.~\ref{fig:TB_band}.}
    \label{fig:TB_phase}
\end{figure}

\begin{figure}[htbp]
    \centering
    \includegraphics[width=0.8\textwidth]{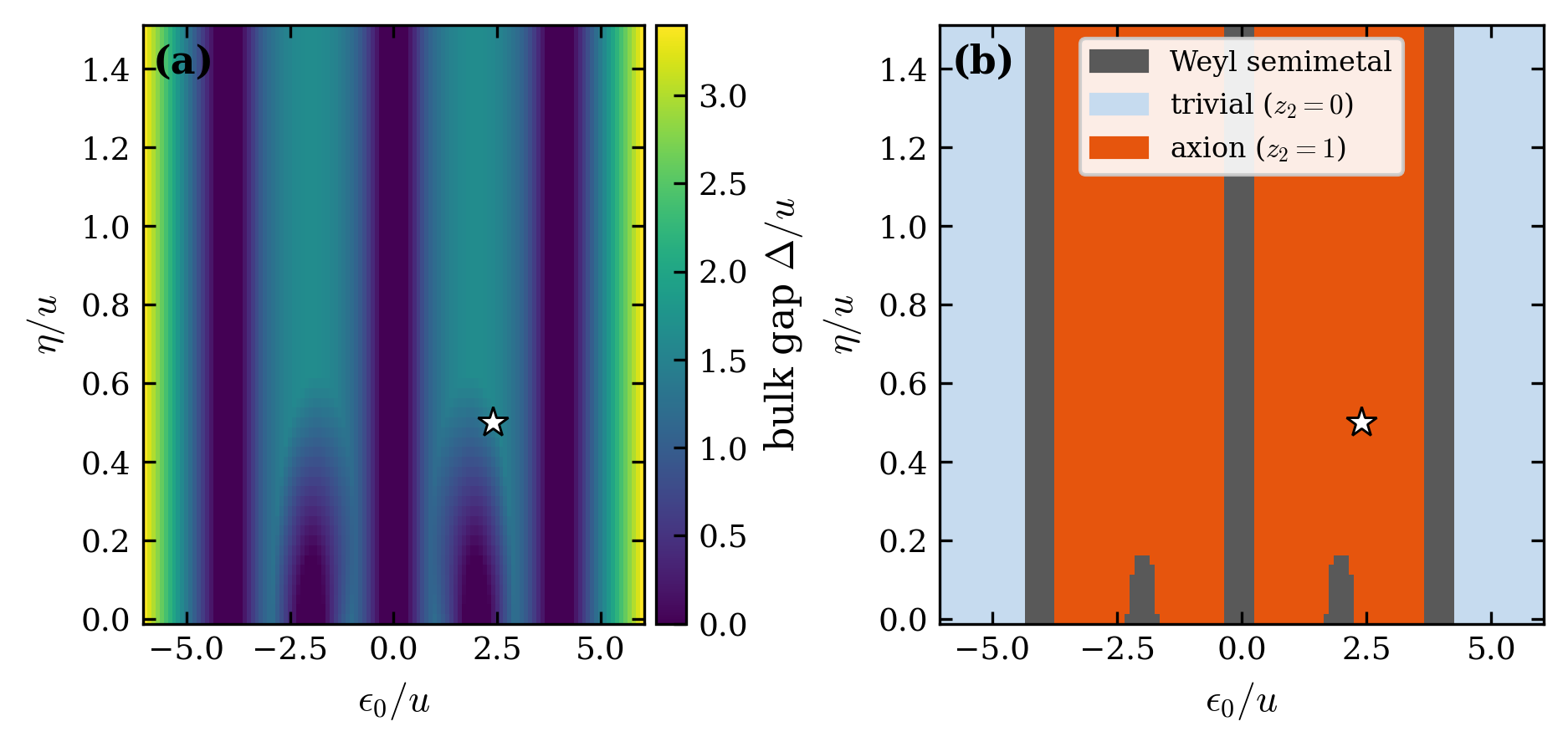}
    \caption{Phase diagram of the model in the $(\epsilon_0,\eta)$ plane for $m_z=0.3$, $\kappa=0.5$: (a) bulk gap $E_3-E_2$ and (b) phases determined from the symmetry indicators. The star marks the parameters of Fig.~\ref{fig:TB_band}.}
    \label{fig:TB_phase_eta}
\end{figure}

We first plot the band gaps and the corresponding topological phase diagrams obtained from the $z_2$ symmetry indicator.
The closed-form spectrum makes the phase diagram exact.
The gap $E_3-E_2=2\min\sqrt{(\rho-m_z)^2+\sin^2k_x+\sin^2k_y}$ can close only on the four $C_2$-invariant lines $(k_x,k_y)\in\{(0,0),(\pi,\pi),(\pi,0),(0,\pi)\}$, and there only where $\rho(k_z)=m_z$.
For small $m_z$ the zero-energy set consists of isolated points, so the gapless regions are Weyl-semimetal wedges around the band inversions at $\epsilon_0=\pm m_z$ ($\Gamma$ and $A$), $4\pm m_z$ ($Z$), and $-4\pm m_z$ ($M$).
The band inversions on the $X$--$R$ line at $\epsilon_0=\pm2$ do not close the gap as long as $m_z<2\eta$, because there $\rho\geq|g|=2\eta$; for larger $m_z$ they open an additional gapless region that terminates the axion phase.
Figure~\ref{fig:TB_phase} shows the bulk gap and the resulting phases in the $(\epsilon_0,m_z)$ plane: a trivial insulator for $|\epsilon_0|>4+m_z$, an axion insulator with $z_2=1$ for $m_z<|\epsilon_0|<4-m_z$, and a Weyl semimetal in between.
Every insulating point has $z_{4S}=\delta_{2S}=0$ and vanishing plane Chern numbers.
The phase diagram in the $(\epsilon_0,\eta)$ plane is additionally plotted in Fig.~\ref{fig:TB_phase_eta}.


\begin{figure}[htbp]
    \centering
    \includegraphics[width=\textwidth]{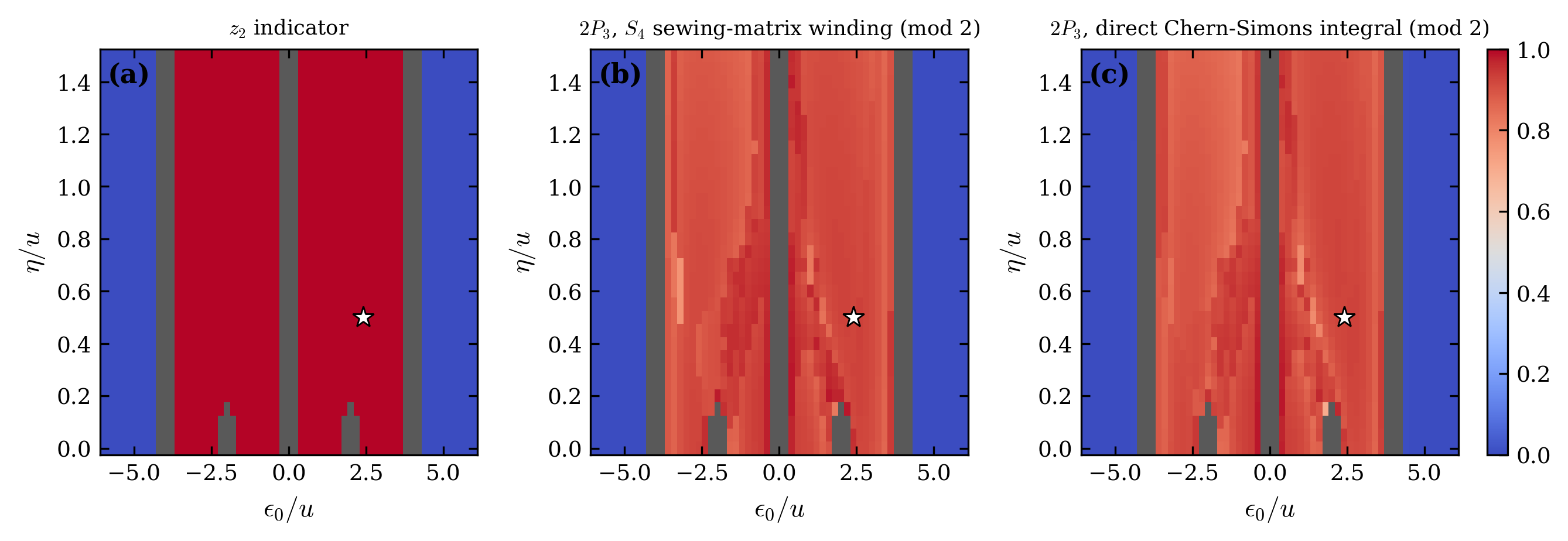}
    \caption{Same as Fig.~\ref{fig:TB_p3map} in the $(\epsilon_0,\eta)$ plane for $m_z=0.3$, $\kappa=0.5$. All 1592 insulating points are resolved and agree with $z_2$.}
    \label{fig:TB_p3map_eta}
\end{figure}

$2P_3$ was also evaluated at every insulating point on a $61\times31$ grid for each phase diagram mentioned above, using both the sewing-matrix mapping degree in Eq.~\eqref{eq:P3_with_B} and the direct Chern-Simons integral in Eq.~\eqref{P3_def}.
The $k$-grid starts at $n_k=24$ and is refined up to $n_k=64$ wherever the smooth frame changes rapidly between neighbouring grid points, which happens only next to the phase boundaries.
The comparison in the $(\epsilon_0,m_z)$ plane is shown in Fig.~\ref{fig:TB_p3map} in Sec.~\ref{sec:tight_binding}.
The same comparison in the $(\epsilon_0,\eta)$ plane, Fig.~\ref{fig:TB_p3map_eta}, contains 1592 insulating points, all of which are resolved and agree with $z_2$.
The $\eta$ dependence is instructive: the axion phase persists down to $\eta\to0$, where inversion and $C_4$ are restored, and the value of $2P_3$ does not change when these additional symmetries are removed, in accordance with the claim that $S_4$ alone quantizes $\theta$.

\bibliographystyle{unsrt}
\bibliography{reference}

\end{document}